\documentclass[Journal]{IEEEtran}
\usepackage[T1]{fontenc}
\usepackage[latin9]{inputenc}
\usepackage{color}
\usepackage{float}
\usepackage{textcomp}
\usepackage{bm}
\usepackage{amsmath}
\usepackage{amssymb}
\usepackage{graphicx}
\usepackage[unicode=true,
 bookmarks=true,bookmarksnumbered=true,bookmarksopen=true,bookmarksopenlevel=1,
 breaklinks=false,pdfborder={0 0 0},pdfborderstyle={},backref=false,colorlinks=false]
 {hyperref}
\hypersetup{pdftitle={Your Title},
 pdfauthor={Your Name},
 pdfpagelayout=OneColumn, pdfnewwindow=true, pdfstartview=XYZ, plainpages=false}

\makeatletter

\providecommand{\tabularnewline}{\\}
\floatstyle{ruled}
\newfloat{algorithm}{tbp}{loa}
\providecommand{\algorithmname}{Algorithm}
\floatname{algorithm}{\protect\algorithmname}

\usepackage[caption=false,font=footnotesize]{subfig}
\usepackage{bm}

\usepackage{cite}
\usepackage[T1]{fontenc}
\usepackage{algorithm}
\usepackage{algorithmic}

\IEEEoverridecommandlockouts

\@ifundefined{showcaptionsetup}{}{%
 \PassOptionsToPackage{caption=false}{subfig}}
\usepackage{subfig}
\makeatother

\begin{document}
\title{Meta-Reinforcement Learning for Timely and Energy-efficient Data Collection
in Solar-powered UAV-assisted IoT Networks}
\author{Mengjie~Yi, Xijun~Wang, Juan Liu, Yan~Zhang, and Ronghui Hou\thanks{M. Yi and R. Hou are with School of Cyber Engineering, Xidian University,
Xi\textquoteright an 710071, China (e-mail: mjyi@stu.xidian.edu.cn,
rhhou@xidian.edu.cn). X.~Wang is with School of Electronics and Information
Technology, Sun Yat-sen University, Guangzhou, 510006, China (e-mail:
wangxijun@mail.sysu.edu.cn). J. Liu is with School of Electrical Engineering
and Computer Science, Ningbo University, Zhejiang, 315211, China (e-mail:
eeliujuan@gmail.com). Y. Zhang is with State Key Lab of Integrated
Service Networks, Information Science Institute, Xidian University,
Xi\textquoteright an, Shaanxi, 710071, China (e-mail: yanzhang@xidian.edu.cn).}}
\maketitle
\begin{abstract}
Unmanned aerial vehicles (UAVs) have the potential to greatly aid
Internet of Things (IoT) networks in mission-critical data collection,
thanks to their flexibility and cost-effectiveness. However, challenges
arise due to the UAV's limited onboard energy and the unpredictable
status updates from sensor nodes (SNs), which impact the freshness
of collected data. In this paper, we investigate the energy-efficient
and timely data collection in IoT networks through the use of a solar-powered
UAV. Each SN generates status updates at stochastic intervals, while
the UAV collects and subsequently transmits these status updates to
a central data center. Furthermore, the UAV harnesses solar energy
from the environment to maintain its energy level above a predetermined
threshold. To minimize both the average age of information (AoI) for
SNs and the energy consumption of the UAV, we jointly optimize the
UAV trajectory, SN scheduling, and offloading strategy. Then, we
formulate this problem as a Markov decision process (MDP) and propose
a meta-reinforcement learning algorithm to enhance the generalization
capability. Specifically, the compound-action deep reinforcement learning
(CADRL) algorithm is proposed to handle the discrete decisions related
to SN scheduling and the UAV's offloading policy, as well as the continuous
control of UAV flight. Moreover, we incorporate meta-learning into
CADRL to improve the adaptability of the learned policy to new tasks.
To validate the effectiveness of our proposed algorithms, we conduct
extensive simulations and demonstrate their superiority over other
baseline algorithms.
\end{abstract}


\IEEEpeerreviewmaketitle{}

\section{Introduction}

In recent years, the Internet of Things (IoT) has rapidly evolved
with the unprecedented popularity of mobile devices, including smart
home appliances and personal devices, to support all aspects of our
daily lives \cite{AI-Fuqaha_IoT-survey_2015IEEECST}. However, the
expanding scale of the IoT poses several challenges for network operators
and service providers. One of the key challenges stems from the fact
that IoT devices typically operate with limited energy resources.
These constraints restrict their capacity to transmit data over long
distances, thus hindering their ability to establish robust connections
with remote base stations (BSs) \cite{Hossein_UAV-IoT-services-survey_IEEEIoT2016}.
Additionally, it is of significant importance to deploy IoT applications
such as environmental sensing and disaster monitoring in challenging
environments, including swamps, deserts, hazardous battlefields, and
contaminated sites \cite{Wu_commu-traj-design-UAV-wireless-net_IEEEWC_2019}.
Unfortunately, in such scenarios, IoT devices are often placed in
locations that lack proper wireless infrastructure. 

Unmanned aerial vehicles (UAVs) are poised to become a pivotal element
of future wireless networks, primarily due to their complete mobility
control, economical operation, and rapid deployment capabilities.
Within the context of IoT networks,  UAVs can adapt their flight paths
to collect environmental information from sensor nodes (SNs) in their
vicinity. Through the establishment of a line-of-sight (LoS) communication
link, UAVs can communicate reliably with SNs, thereby contributing
to a reduction in the transmit power requirements of SNs and an extension
of the network's operational lifespan. However, this reduction in
transmission energy for SNs comes at the expense of consuming the
propulsion energy of the UAV itself \cite{Yang_energy-tradeoff-G2A-traj-design_IEEETVT2018}.
This presents a challenge in the deployment of UAVs for data collection
from SNs in IoT networks, given the finite onboard energy of UAVs.
While the optimization of energy consumption can extend the UAV's
operational lifetime to a certain extent \cite{Abedin_Data-fresh-EE-drl_IEEETITS2021,RDing_3DTra-Freq-EE-FairC-DRL_2020TWC},
it cannot completely overcome the inherent energy limitation problem.
Consequently, efficient energy replenishment mechanisms are imperative
for the practical deployment of UAVs. Solar-powered UAVs, in particular,
have attracted considerable interest for their ability to sustain
flight operations over extended periods \cite{Sun_Solar-power-UAV_TCOM2019,Fu_Solar-power-uav_TVT2021}.
Importantly, these solar-powered UAVs eliminate the necessity for
periodic returns to a charging station to replenish their energy reserves
\cite{UAV_BS_wirelessCharge2018}. This feature not only enhances
their operational autonomy but also holds promise for a wide range
of applications, where uninterrupted aerial data collection or surveillance
is essential.

Numerous studies have concentrated on UAV-assisted data collection
within IoT networks, with a primary focus on traditional performance
metrics like throughput and latency as their principal optimization
objectives \cite{Qian_UAV-MEC_2023IEEEIoTJ,Wang_Coverage-UAVs_IEEETWC2022,Zhang_UAVs-BSs-IEEETWC2023}.
Nonetheless, these conventional metrics may not fully reflect the
timeliness of the data. Fortunately, the emergence of the age of information
(AoI) has ushered in a novel perspective for assessing the freshness
of data from SNs at the receiver's perspective \cite{S.Kaul_Mini_AoI_VehicularNet,WXJ_AoI-Vehicular_IEEENetwork2022}.
Many efforts have employed deep reinforcement learning (DRL) algorithms
to refine system parameters, such as UAV trajectories and SN associations,
with the goal of reducing the AoI in UAV-assisted IoT networks \cite{abd-elmagidDeepReinforcementLearning2019,Zhang_traj-UAVs-BSs-GNN-MARL_2023TWC,Chu_SpeedControlEnergyReplenishSTDRL_2022IoT}.
However, a noteworthy limitation of policies obtained through DRL
algorithms is their limited generalization capability. When the task
undergoes changes, the policies previously learned often become less
applicable, resulting in a decline in performance.

Driven by these challenges, we investigate a issue of timely and energy-efficient
data collection assisted by a solar-powered UAV in IoT networks.
Within the system under consideration, each SN stochastically samples
the environment and generates update packets. The update packets are
subsequently gathered by a UAV, which caches them within its onboard
buffer. When the UAV reaches suitable locations, it offloads the gathered
data to the data center (DC). We employ a DRL approach to achieve
the dual objectives of reducing the average AoI of SNs and conserving
the energy consumption of the UAV. Furthermore, we augment DRL with
meta-learning to enhance the algorithm's capacity for generalization.
The primary contributions of this paper can be outlined as follows:
\begin{itemize}
\item We study the timely and energy-efficient data collection problem in
solar-powered UAV-assisted IoT networks. We undertake the joint optimization
of the UAV's velocity, SN scheduling, and data offloading decisions
while considering the energy and kinematic constraints inherent to
the UAV. This problem is formulated as a Markov decision process (MDP).
Given the combination of continuous and discrete actions in this problem,
we design a compound-action DRL (CADRL) algorithm to minimize the
weighted sum of the average AoI of SNs and the energy consumption
of the UAV.
\item To enhance the CADRL algorithm's ability to adapt to new tasks and
improve its generalization, we propose a meta-learning-based CADRL
(MLCADRL) algorithm. MLCADRL is equipped to acquire a meta-policy
from a multitude of learning tasks and thus converge quickly on new
tasks, resulting in increased scalability and adaptability.
\item We undertake comprehensive simulations to assess the performance of
the algorithms we propose. Our results reveal that the CADRL-based
algorithm adeptly coordinates the UAV's velocity, the scheduling of
SNs, and the offloading strategy, ultimately surpassing baseline algorithms
by achieving the lowest system cost. Moreover, the MLCADRL-based algorithm
proves effective in accelerating convergence when dealing with new
tasks.
\end{itemize}

The rest of this paper follows this structure: Section \ref{sec:Related-Work}
provides an overview of the related research. Section \ref{sec:System-Model-Problem-Formulation}
presents the system model and problem formulation. In Section \ref{sec:Meta-RL},
CADRL-based and MLCADRL-based data collection algorithms are proposed.
The simulation results are presented in Section \ref{sec:Simulation-Results}.
Finally, conclusions are presented in Section \ref{sec:Conclusions}.

\section{Related Work\label{sec:Related-Work}}

\subsection{UAV-assisted Data Collection in IoT Networks}

In the realm of UAV-assisted IoT networks, various studies  have
been dedicated to optimizing data collection with a focus on ensuring
the freshness of information. Zhu et al. \cite{Zhu_UAV-traj-AoI-Transformer_TWC2023}
employed transformers and weighted A{*} to optimize  UAV hover points
for  cumulative AoI minimization. Liu et al. \cite{JLiu_UAV_AoIWSN}
centered on the optimization of UAV trajectory and SN associations,
and proposed two optimization problems aiming at minimizing the maximum
and average AoI, respectively. Furthermore, the work by \cite{Liu_UAV-traj-AoI-const-env-monitor_IoTJ2022}
considered the UAV's onboard energy, focusing on minimizing the completion
time of the UAV's mission while considering AoI and UAV onboard energy
constraints. The aforementioned studies primarily explored scenarios
involving a single UAV. In contrast, other studies introduced the
use of multiple UAVs \cite{Long_AoIUAVs_2022GLOBECOM,Zhang_AoI-UAV-mURLLC-FBC_JSAC2021,WXJ_MultiUAVs-AoI_TCOM2023,Liu_AoI-task-assign-traj-multiUAVs_IoTJ2022}.
Long et al. \cite{Long_AoIUAVs_2022GLOBECOM} investigated the use
of multiple UAVs for transmitting data from ground users to distant
BSs. They adopted a multi-stage stochastic optimization approach to
reduce the long-term AoI through trajectory and scheduling optimization.
Additionally, Zhang et al. \cite{Zhang_AoI-UAV-mURLLC-FBC_JSAC2021}
studied the AoI minimization in large-scale, ultra-reliable, and low-latency
communication scenarios, considering statistical delay and bit error
rate constraints. Wang et al. \cite{WXJ_MultiUAVs-AoI_TCOM2023} used
a multi-agent DRL approach to jointly optimize trajectories and SN
scheduling of UAVs to minimize AoI. Furthermore, Liu et al. \cite{Liu_AoI-task-assign-traj-multiUAVs_IoTJ2022}
examined a system involving multiple UAVs taking off from a DC, collecting
data from ground SNs, distributing it to users, and then returning
to the DC. They minimized AoI by optimizing task assignment, interaction
point selection, and UAV flight trajectories. 

In the above studies, a common assumption was made that the SNs followed
a generate-at-will generation policy \cite{Sun_how-keep-data-fresh_IEEETIT2017,MAAE_DRL_AoI_RFPower_2020}.
However, it is worth noting that in certain scenarios, SNs may utilize
their own independent sampling mechanisms, separate from the data
transmission scheme. Since both the sampling and transmission processes
can influence the AoI, it becomes imperative to jointly consider these
two factors to effectively reduce the AoI. Recognizing the uncertainty
surrounding the SN's sampling process, Zhou et al. \cite{Zhou_DRL-AoI-UAV-random-sample_WCSP2019}
utilized a DQN-based algorithm to optimize the UAV's trajectory with
the goal of minimizing the AoI. Tong et al. \cite{Tong_DRL-EE-AoI-uav_ICC2020}
assumed that SNs can sample information at fixed or random rates.
Their optimization was the UAV's flight trajectory, with the dual
objective of minimizing the Age of Information (AoI) and reducing
data packet loss rates. Building upon the work of \cite{Tong_DRL-EE-AoI-uav_ICC2020},
Li et al. \cite{Li_VDN-uav-aoi_IoTJ2023} further extended their investigation
to scenarios involving multiple UAVs.

The aforementioned studies predominantly dealt with battery-powered
UAVs, which inherently possess limited operating durations due to
their constrained energy storage capacity. In contrast, solar-powered
UAVs offer a promising avenue for extending operational periods by
harnessing solar energy from the sun  \cite{Sun_Solar-power-UAV_TCOM2019,Fu_Solar-power-uav_TVT2021}.
 Sami et al. \cite{Sam_CDRL-solarUAV-noma_IEEEJSAC2021} focused on
optimizing UAV altitude adjustments and channel access management.
They achieved this through the use of  constrained DRL algorithms,
with the primary objective of maximizing network capacity. Furthermore,
Zhang et al. \cite{Zhang_DRL-hybrid-powered-uav_IEEEIoTJ2023} extended
this optimization by considering the three-dimensional trajectory
and time allocation of solar-powered UAVs. This extension aimed to
enhance overall network throughput while accommodating various constraints,
including energy constraints, quality of service demands, and shifting
flight conditions. \cite{Fu_Solar-power-uav_TVT2021}  detailed the
application of solar-powered UAVs in the data collection process from
IoT devices and simultaneously recharged these devices using laser
technology. The primary aim was to optimize the UAV's energy resources
while satisfying the requirements of IoT devices. Moreover, in \cite{Zhang_EE-solar-powered-UAV_IEEETMC2022},
the UAV could acquire energy from both solar power and charging stations.
Through optimization of the UAV's trajectory, a delicate balance was
struck between achieving average data transmission rates, minimizing
total energy consumption, and ensuring fair coverage of IoT terminals.
While these studies explored various aspects of solar-powered UAV
operations in IoT networks, none of them specifically addressed the
optimization of AoI. 

\subsection{ DRL Methods in UAV-assisted Data Collection}

Over the past few years, DRL has received significant attention due
to its impressive success in intricate tasks such as international
Go, games, and controlling complex machinery for various operations
\cite{silver2018general,berner2019dota,rajeswaran2017learning}. This
success has propelled the application of DRL methods in UAV-assisted
data collection for IoT networks \cite{Wang_Coverage-UAVs_IEEETWC2022,Zhou_DRL-AoI-UAV-random-sample_WCSP2019,Tong_DRL-EE-AoI-uav_ICC2020,Li_VDN-uav-aoi_IoTJ2023,Sam_CDRL-solarUAV-noma_IEEEJSAC2021}.
Nevertheless, the action space considered in these studies is typically
confined to either continuous or discrete domains \cite{Yuan_AC-energy-mini-UAV_TVT2021,Liu_RL-Mul-UAV-deploy-move_TVT2019,Seid_blockchain-energy-harvesting-UAVs-MADRL_JSAC2022,Sun_AoI-TD3-UAV_IoTJ2021,LK_JointFlightCruiseControlDataCollectUAVaidedIoT:DRL_IoT2021}.
Applying these DRL algorithms directly to issues involving both discrete
and continuous action spaces is not feasible. To overcome this obstacle,
Hu et al. \cite{Hu_CooperativeUAV-MADRL-CA2C_TCOM2020} studied UAVs
engaged in collaborative perception and transmission for sensing tasks.
They proposed the compound action actor-critic (CA2C) method to optimize
UAV trajectories and task selection for AoI reduction. Akbari et al.
\cite{Akbari_AoI-VNF-compound-action_JSAC2021} investigated the optimization
problem of virtual network function placement and scheduling in industrial
IoT networks and also utilized the CA2C method to address the problem
with compound actions. The CA2C algorithm can be seen as a combination
of DQN and DDPG, in which the continuous actions selected by DDPG
serve as inputs to DQN. This may increase the sum of Q values at the
cost of decreasing the maximum Q value \cite{fan2019hybrid}. In
addition, the DRL approaches proposed in \cite{Hu_CooperativeUAV-MADRL-CA2C_TCOM2020,Akbari_AoI-VNF-compound-action_JSAC2021}
fail to extract well-generalized knowledge from training tasks, resulting
in the need for DRL agents to start training anew when tackling new
tasks.

There has been some work exploring how to improve the generalizability
of DRL algorithms in UAV-assisted IoT data collection scenarios. Zhu
et al. \cite{Zhu_UAV-data-collection-S2S-DRL-IEEEIoTJ2022} investigated
the minimization of total energy consumption through the joint design
of the UAV's trajectory and cluster head selection. They proposed
a DRL method with a sequential model to address this issue and validated
the algorithm's generalization capability. However, combining a sequence-to-sequence
model with DRL can present challenges because of the intricate training
process. Chu et al. \cite{Chu_SpeedControlEnergyReplenishSTDRL_2022IoT}
aimed at maximizing data collection capacity and energy efficiency,
and proposed a transfer learning combined DRL algorithm to enhance
the efficacy of the DRL-based algorithm on new tasks. Since single-task
transfer learning may suffer from learning bias, Yi et al. \cite{Yi_Multi-task-DRL-UAV-IEEEIoTJ2023}
utilized multi-task transfer learning in conjunction with DRL for
the purpose of optimizing the UAV's trajectory and recharging decisions
that lead to the reduction of the AoI. While multi-task transfer DRL
can leverage knowledge acquired from training tasks to enhance its
performance on new tasks,  it exhibits relatively longer convergence
times when compared to meta DRL approaches. Lu et al. \cite{Lu_UAV-metaDRL-traj_INFOCOM2023}
explored UAV data collection from ground nodes, with a focus on maximizing
data collected by optimizing the UAV's trajectory. A meta-DRL-based
algorithm was proposed to enhance  generalization capabilities. However,
this work focused on UAV actions limited to a single type of discrete
action, specifically, flight direction.

\section{System Model and Problem Formulation\label{sec:System-Model-Problem-Formulation}}

\begin{figure}[tb]
\centering

\includegraphics[width=0.4\textwidth]{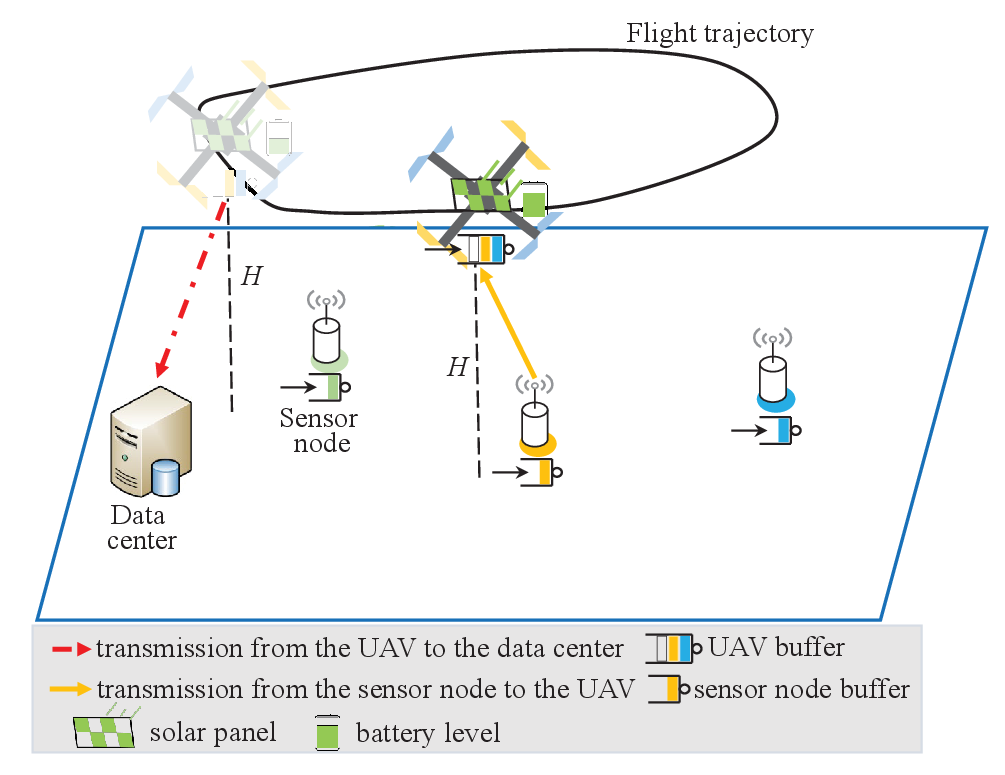}\caption{The scene of solar-powered UAV-assisted data collection. The UAV collects
status updates from SNs and caches them in its onboard buffer. During
the collection process, the UAV offloads the cached data to the DC.
Additionally, the UAV can harness energy from the environment through
the solar panel.  \label{fig:system_model}}
\end{figure}

 As depicted in Figure \ref{fig:system_model}, we consider solar-powered
UAV-assisted data collection within the IoT network, in which $N$
sensor nodes (SNs) are randomly distributed across the area. Each
SN performs random environmental sampling, and the status update for
each SN follows a Poisson process with an arrival rate of $\lambda_{0}$.
The collected data is cached in the buffer as a packet, each containing
$w$ bits and accompanied by a timestamp.  We use $\mathcal{N}=\{1,2,\ldots,N\}$
to denote the set of all the SNs, with the location of SN $n\in\mathcal{N}$
specified as $\boldsymbol{C}_{n}=(x_{n},y_{n})$. The location of
the DC is represented by $\boldsymbol{C}_{0}=(x_{0},y_{0})$. We define
$\mathcal{N}^{+}=\{0\}\cup\mathcal{N}$, signifying the set of all
nodes, including the DC.  

We consider a discrete-time system where time is partitioned into
equidistant time slots. Each slot has a duration of $\tau_{0}$ seconds.
Suppose that the UAV is required to enable data collection for $T$
slots. Specifically, the rotary-wing UAV serves as a mobile relay,
taking off from the DC, flying over SNs to collect data packets, and
temporarily storing the collected data packets in its onboard buffer.
Once the UAV reaches an appropriate location, it offloads all buffered
data packets to the DC for processing.\textcolor{black}{{} It is assumed
that the UAV operates at a consistent altitude, denoted by $H$. Thus,
the UAV's location can be denoted by its ground-level projection at
time slot $t$,} $\boldsymbol{C}_{u}(t)=[x_{u}(t),y_{u}(t)]$\textcolor{black}{.
At the beginning of time slot $t$, the UAV's velocity can be represented
using polar coordinates and denoted by $\boldsymbol{v}(t)=(v_{s}(t),\phi(t))$.
Here, $v_{s}(t)=\left\Vert \boldsymbol{v}(t)\right\Vert $ signifies
the UAV's speed, while $\phi(t)$ stands for the velocity direction,
within the range $0\leq\phi(t)\leq2\pi$. We consider a consistent
acceleration $\bm{a}_{\textrm{c}}(t)=\dot{\boldsymbol{v}}(t)$ throughout
a time slot, leading to the update of the velocity as $\bm{v}(t+1)=\bm{v}(t)+\bm{a}_{\textrm{c}}(t)\tau_{0}$.
We assume that the }rotary-wing\textcolor{black}{{} UAV can swiftly
alter its orientation as a time slot commences and then remain that
way for the duration of the time slot, since the }rotary-wing\textcolor{black}{{}
UAV can readily turn by modifying the rotation of its rotors. In operations,
the UAV faces kinematic restrictions. Specifically, the UAV's speed
at slot $t$ does not go beyond its maximum limit $v_{s}^{\textrm{max}}$,
i.e., $v_{s}(t)\leq v_{s}^{\textrm{max}}$, and the UAV's turning
angle at slot $t$, $\triangle\phi(t)=\phi(t)-\phi(t-1)$, does not
go beyond its maximum limit $\triangle\phi_{\max}$, i.e., $\mid\triangle\phi(t)\mid\leq\triangle\phi_{\max}$.
Let $\boldsymbol{V}=(\boldsymbol{v}(1),\boldsymbol{v}(2),\ldots,\boldsymbol{v}(T))$
denote the series of UAV velocities. The UAV's flight path encompasses
a series of locations it crosses, i.e., }$\bm{p}=(\boldsymbol{C}_{u}(1),\boldsymbol{C}_{u}(2),\ldots,\boldsymbol{C}_{u}(T))$,
where $\boldsymbol{C}_{u}(1)=\boldsymbol{C}_{0}$.

During each time slot, the UAV is required to make a determination
regarding the scheduling of a specific SN. The UAV's SN scheduling
vector is denoted as $\boldsymbol{b}=(b(1),b(2),\ldots,b(T))$, with
$b(t)=n$ indicating that SN $n$ will transmit its status updates
to the UAV during time slot $t$, while $b(t)=0$ signifies that no
SN is scheduled for time slot $t$.  On the other hand, the UAV needs
to determine at each time slot whether to offload data packets in
its buffer to the DC based on the SNs' AoI, the UAV's energy level,
and the link status between the UAV and the DC. We use $\boldsymbol{q}=(q(1),q(2),\ldots,q(T))$
to represent the UAV's offloading vector, with $q(t)\in\{0,1\}$.
Specifically, $q(t)=1$ means that the data packets in the UAV's buffer
are offloaded to the DC in time slot $t$; otherwise, $q(t)=0$. Note
that in a given time slot, the UAV cannot simultaneously schedule
SN data transmission and offload data to the DC. It can first schedule
a SN for a duration of $\tau_{\textrm{s}}$ and then offload data
to the DC within a duration of $\tau_{0}-\tau_{\textrm{s}}$. 

The UAV's maximum onboard energy is $E_{\textrm{max}}$, and its energy
level in slot $t$ is indicated by $E(t)\in[0,E_{\textrm{max}}]$.
To ensure that the UAV does not crash due to insufficient energy,
we assume that the UAV has two modes, namely, working and restoring,
which are determined by the UAV's energy level. Let $m(t)\in\{0,1\}$
denote the mode of the UAV. $m(t)=0$ means that the UAV is in the
restoring mode for the time slot $t$; otherwise, the UAV is in the
working mode. In the working mode, the UAV can gather status updates
from SNs and offload the data packets in its buffer to the DC. When
the UAV's energy level falls below the threshold $E_{\textrm{th}}^{1}$,
i.e., $E(t)<E_{\textrm{th}}^{1}$, it switches to the restoring mode.
The UAV lands on the ground and waits for the harvested energy. When
the UAV's energy level exceeds a certain threshold $E_{\textrm{th}}^{2}$,
i.e., $E(t)\geq E_{\textrm{th}}^{2}$, it switches to the working
mode, ascends from the ground to altitude $H$, and continues its
data collection task.

\subsection{Transmission Model}

In this work, we account for both large-scale and small-scale fading
in the channel model \cite{Samir_large&small_scale_TWC2020,Tran_large&small_scale_TWC2022}.
Let $h_{u,n}(t)$ represent the channel coefficient between the UAV
and node $n\in\mathcal{N}^{+}$ in time slot $t$, as expressed by
\begin{equation}
h_{u,n}(t)=\sqrt{g_{u,n}(t)}\widetilde{h}_{u,n}(t),\label{eq:channel_gain}
\end{equation}
where $g_{u,n}(t)$ denotes the large-scale fading factor and $\widetilde{h}_{u,n}(t)$
is the small-scale attenuation coefficient with $\mathbb{E}[|\widetilde{h}_{u,n}(t)|^{2}]=1$.
The large-scale channel fading is either line-of-sight (LoS) or non-line-of-sight
(NLoS), depending on the propagation environment. Thus, we assume
that the UAV-node $n\in\mathcal{N}^{+}$ channel experiences LoS fading
in a probabilistic manner. The LoS probability between the UAV and
node $n\in\mathcal{N}^{+}$ is expressed as \cite{al2014optimal}
\begin{equation}
p_{u,n}^{\text{LoS}}(t)=\frac{1}{1+\beta\exp\left(\text{\textminus}\beta'\left(\frac{180}{\pi}\arcsin\left(\frac{H}{d_{u,n}(t)}\right)\text{\textminus}\beta\right)\right)},
\end{equation}
where $\beta$ and $\beta'$ are constants that depend on the environment,
and $d_{u,n}(t)=\sqrt{H^{2}+\Vert\boldsymbol{C}_{u}(t)-\boldsymbol{C}_{n}\Vert^{2}}$
signifies the Euclidean distance between the UAV and node $n\in\mathcal{N}^{+}$
in slot $t$. Thus, the large-scale channel gain between the UAV and
node $n\in\mathcal{N}^{+}$ in slot $t$ can be represented as
\begin{equation}
g_{u,n}(t)=\begin{cases}
\beta_{0}[d_{u,n}(t)]^{-\varsigma}, & \textrm{w.p. \ensuremath{p_{u,n}^{\text{LoS}}(t)}},\\
\kappa\beta_{0}[d_{u,n}(t)]^{-\varsigma}, & \textrm{w.p. }\ensuremath{1-p_{u,n}^{\text{LoS}}(t)}\textrm{,}
\end{cases}\label{eq:large_scale_channel-gain}
\end{equation}
where $\beta_{0}$ represents the channel gain at a reference distance
of one meter, $\varsigma$ denotes the path-loss exponent, $\kappa$
$(\kappa<1)$ denotes the additional attenuation factor associated
with NLoS, and ``w.p.'' is an abbreviation for ``with probability''.

The signal-to-noise ratio (SNR) of the air-to-ground (A2G) channel
linking the UAV and SN $n\in\mathcal{N}$ in time slot $t$ can be
described as
\begin{align}
\xi_{u,n}(t) & =\frac{P_{s}|h_{u,n}(t)|^{2}}{\sigma^{2}},\label{eq:SNR}
\end{align}
where $P_{s}$ denotes the transmission power of each SN, and $\sigma^{2}$
denotes the noise power of the A2G channel. If $\xi_{u,n}(t)$ is
not lower than the threshold $\xi_{\textrm{th}}$, i.e., $\xi_{u,n}(t)\geq\xi_{\textrm{th}}$,
the UAV successfully receives the status updates from SN $n$ during
time slot $t$. Otherwise, it fails.

The transmission rate of the A2G channel that links the UAV and the
DC in time slot $t$ is given by
\begin{equation}
R_{u}(t)=B\log_{2}(1+\frac{P_{\textrm{c}}^{\textrm{t}}|h_{u,0}(t)|^{2}}{\sigma^{2}}),\label{eq:Trans_rate}
\end{equation}
where $P_{\textrm{c}}^{\textrm{t}}$ denotes the UAV's transmission
power and $B$ denotes the channel bandwidth. When the UAV offloads
the data packets in its buffer to the DC within a specified time $\tau(t)$
and $\frac{W(t)}{R_{u}(t)}\leq\tau(t)$, it is considered that the
data offloading is successful, where $W(t)$ is the size of data in
the UAV's buffer at slot $t$. Otherwise, it fails. Specifically,
if the UAV only offloads data to the DC during the time slot $t$,
then $\tau(t)=\tau_{0}$. If the UAV both schedules SNs and offloads
data during time slot $t$, then $\tau(t)=\tau_{0}-\tau_{\textrm{s}}$.

\subsection{Age of Information}

SN $n\in\mathcal{N}$ utilizes a random sampling strategy to sample
information from its surroundings and packages it into a time-stamped
data packet. The data packet is placed in the buffer of SN $n$.
Let $k_{n}(t)\in\{0,1\}$ denote the packet arrival status of SN $n$
at time slot $t$. $k_{n}(t)=1$ means SN $n$ samples its surrounding
environment, and a new data packet reaches SN $n$ in time slot $t$.
Otherwise, $k_{n}(t)=0$. When the buffer of SN $n$ contains data
and a new data packet arrives, the old data packet will be replaced
by the new one. 

Let $Y_{n}(t)$ keep track of the lifetime of the data packet in SN
$n$ if there exits one at time slot $t$. The update of $Y_{n}(t)$
is 
\begin{equation}
Y_{n}(t)=\begin{cases}
0, & \textrm{if }k_{n}(t)=1,\\
Y_{n}(t-1)+1, & \textrm{otherwise}.
\end{cases}\label{eq:SN_life-1}
\end{equation}
 In (\ref{eq:SN_life-1}), if a new packet arrives at SN $n$ at
time slot $t$, i.e., $k_{n}(t)=1$, $Y_{n}(t)$ is set to zero. Otherwise,
$Y_{n}(t-1)$ is incremented by one.

 The inherent unpredictability of the channel connecting SN $n$
and the UAV implies that, even with scheduling, the transmission may
fail to be successful.  Let $z_{n}(t)\in\{0,1\}$ denote the uploading
status of SN $n$. Specifically, $z_{n}(t)=1$ signifies the successful
transmission of SN $n$ to the UAV, i.e., $b(t)=n\textrm{ and }\xi_{u,n}(t)\geq\xi_{\textrm{th}}$;
otherwise, $z_{n}(t)=0$. The UAV has a buffer of capacity $Nw$
that can cache one data packet for each SN. When a data packet from
SN $n$ already exists in the UAV's buffer and a new data packet from
SN $n$ is received, the old data packet of SN $n$ will be replaced
with this new data packet. $U_{n}(t)$ is used to track the lifetime
of the data packet of SN $n$ in the UAV's buffer if there exists
one. Therefore, the update of $U_{n}(t)$ is represented as 
\begin{equation}
U_{n}(t)=\begin{cases}
Y_{n}(t)+1, & \textrm{if }z_{n}(t)=1,\\
U_{n}(t-1)+1, & \textrm{otherwise}.
\end{cases}\label{eq:UAV_life1}
\end{equation}
 In (\ref{eq:UAV_life1}), if the update packet from SN $n$ is
effectively delivered to the UAV, i.e., $z_{n}(t)=1$, $U_{n}(t)$
is set to $Y_{n}(t)+1$. Otherwise, $U_{n}(t-1)$ is increased by
one.

 Let $o(t)\in\{0,1\}$ indicate the offload status of the UAV. In
particular, $o(t)=1$ represents that the data packets are successfully
offloaded from the UAV to the DC, i.e., $q(t)=1$ and $\frac{W(t)}{R_{u}(t)}\leq\tau(t)$;
otherwise, $o(t)=0$. The AoI is utilized to depict the freshness
of data packets originating from SNs at the DC. More specifically,
$\delta_{n}(t)$, which represents the AoI of SN $n$, is defined
as the time that has passed since the DC received the latest status
update. This can be expressed as:
\begin{equation}
\delta_{n}(t)=\begin{cases}
U_{n}(t), & \textrm{if }U_{n}(t)\geq0\textrm{ and }o(t)=1,\\
\delta(t-1)+1, & \text{otherwise}.
\end{cases}\label{eq:AoI}
\end{equation}
In (\ref{eq:AoI}), if a data packet of SN $n$ is cached in the UAV's
buffer in time slot $t$, i.e., $U_{n}(t)\geq0$, and the cached data
packets are successfully offloaded by the UAV to the DC within time
slot $t$, i.e., $o(t)=1$, $\delta_{n}(t)$ is set to $U_{n}(t)$.
Otherwise, $\delta_{n}(t-1)$ is increased by one.

\subsection{Energy Model}

\subsubsection{Energy Consumption Model}

The energy consumption of a rotary-wing UAV is chiefly determined
by its propulsion and communication necessities. The following is
an expression for the UAV's propulsion power during time slot $t$\textcolor{black}{{}
\cite{RDing_3DTra-Freq-EE-FairC-DRL_2020TWC}:
\begin{align}
P_{\textrm{c}}^{\textrm{p}}(t)= & n_{r}\Bigg[\frac{\chi}{8}\left(\frac{T_{\textrm{h}}(t)}{x_{T}\rho A}+3\left(v_{s}(t)\right)^{2}\right)\sqrt{\frac{T_{\textrm{h}}(t)\rho x_{s}^{2}A}{x_{T}}}+\nonumber \\
 & \frac{1}{2}d_{0}\rho x_{s}A\left(v_{s}(t)\right)^{3}+(1+x_{f})T_{\textrm{h}}(t)\times\nonumber \\
 & \left(\sqrt{\frac{\left(T_{\textrm{h}}(t)\right)^{2}}{4\rho^{2}A^{2}}+\frac{\left(v_{s}(t)\right)^{4}}{4}}-\frac{\left(v_{s}(t)\right)^{2}}{2}\right)^{\frac{1}{2}}\Bigg],\label{eq:Energy_Consumption}
\end{align}
where $n_{r}$ represents the number of rotors, $x_{T}$, $x_{s}$,
and $x_{f}$ indicate the thrust coefficient based on disc area, rotor
solidity, and the incremental correction factor of induced power,
respectively, $\chi$ is the local blade section drag coefficient,
$\rho$ is air density, $A$ denotes the disc area for each rotor,
$d_{0}$ denotes the fuselage drag ratio for each rotor, and $T_{\textrm{h}}(t)$
indicates the thrust of each rotor. To better elaborate, we only take
into account the acceleration that is parallel to the velocity \cite{RDing_3DTra-Freq-EE-FairC-DRL_2020TWC}.
Hence, each rotor's thrust can be expressed as 
\begin{equation}
T_{\textrm{h}}(t)=\frac{1}{n_{r}}\left[\left(Ma_{\textrm{c}}(t)+\frac{1}{2}\rho\left(v_{s}(t)\right)^{2}S_{FA}\right)^{2}+(Mg)^{2}\right]^{1/2},\label{eq:Thrust_rotor}
\end{equation}
where $a_{\textrm{c}}(t)=(v_{s}(t+1)-v_{s}(t))/\tau_{0}$, $S_{FA}$
denotes the fuselage equivalent flat plate area, $M$ and $g$ are
the UAV's weight and the gravity acceleration, respectively.}

The energy consumption of the UAV $e_{\textrm{c}}(t)$ in a time slot
is discussed in the following four scenarios: If the UAV is in a restoring
mode at time slot $t$, i.e., $m(t)=0$, and the energy level is less
than $E_{\textrm{th}}^{2}$, i.e., $E(t)<E_{\textrm{th}}^{2}$, the
energy consumption is zero; if the UAV is in a working mode at time
slot $t$, i.e., $m(t)=1$, the energy level is greater than $E_{\textrm{th}}^{1}$,
i.e., $E(t)\geq E_{\textrm{th}}^{1}$, and it schedules an SN while
offloading data packets, i.e., $b(t)\neq0$ and $q(t)=1$, the energy
consumption is $\tau_{0}P_{\textrm{c}}^{\textrm{p}}(t)+(\tau_{0}-\tau_{c})P_{\textrm{c}}^{\textrm{t}}$;
if the UAV is in a working mode at time slot $t$, i.e., $m(t)=1$,
the energy level is greater than $E_{\textrm{th}}^{1}$, i.e., $E(t)\geq E_{\textrm{th}}^{1}$,
and it only offloads data packets without scheduling an SN, i.e.,
$b(t)=0$ and $q(t)=1$, the energy consumption is $\tau_{0}\left(P_{\textrm{c}}^{\textrm{p}}(t)+P_{\textrm{c}}^{\textrm{t}}\right)$;
in other cases, the UAV uses its propulsion energy just for flying.
Hence, the UAV's energy consumption $e_{\textrm{c}}(t)$ in time slot
$t$ is summarized in (\ref{eq:e_c(t)}) on the next page.
\begin{figure*}[tb]
\begin{equation}
e_{\textrm{c}}(t)=\begin{cases}
0, & \textrm{if }m(t)=0\textrm{ and }E(t)<E_{\textrm{th}}^{2},\\
\tau_{0}P_{\textrm{c}}^{\textrm{p}}(t)+(\tau_{0}-\tau_{c})P_{\textrm{c}}^{\textrm{t}}, & \textrm{if }m(t)=1,E(t)\geq E_{\textrm{th}}^{1},b(t)\neq0,\textrm{and }q(t)=1,\\
\tau_{0}\left(P_{\textrm{c}}^{\textrm{p}}(t)+P_{\textrm{c}}^{\textrm{t}}\right), & \textrm{if }m(t)=1,E(t)\geq E_{\textrm{th}}^{1},b(t)=0,\textrm{and }q(t)=1,\\
\tau_{0}P_{\textrm{c}}^{\textrm{p}}(t), & \textrm{otherwise}.
\end{cases}\label{eq:e_c(t)}
\end{equation}
\hrulefill
\end{figure*}

\subsubsection{Energy Harvesting Model}

Providing a sustainable energy supply for the UAV with limited battery
capacity is a crucial challenge. Assuming that the UAV can harvest
energy from sunlight, solar energy is not reliable and depends on
various factors, such as weather and altitude. The process of solar
energy harvesting by the UAV is portrayed as an independent Bernoulli
process with parameter $\lambda_{1}$ \cite{sn-energy-bonuli1_JSAC2016}.
 This signifies that the probability of solar energy arriving at the
UAV at each time slot is $\lambda_{1}$. The solar energy reaching
the UAV during time slot $t$ can be expressed as \cite{Fu_Solar-power-uav_TVT2021}
\begin{equation}
e_{\textrm{r}}(t)=\tau_{0}\eta_{1}S_{u}G_{1}\left(\varphi_{1}-\varphi_{2}e^{-\frac{h(t)}{h_{1}}}\right),\label{eq:-5-1}
\end{equation}
where $h(t)$ is the UAV's altitude in time slot $t$, $\eta_{1}$
denotes the efficiency of energy conversion, $S_{u}$ indicates the
area of the solar panel that effectively receives light, $G_{1}$
represents the average solar radiation on the ground, $\varphi_{1}$
and $\varphi_{2}$ are the maximum atmospheric transmittance value
and the atmosphere's extinction coefficient, respectively, and $h_{1}$
denotes the earth's scale height. Therefore, the energy harvested
by the UAV during time slot $t$ is given by
\begin{equation}
e_{\textrm{h}}(t)=\begin{cases}
e_{\textrm{r}}(t), & \textrm{w.p}.\,\lambda_{1},\\
0, & \textrm{w.p.\,1-\ensuremath{\lambda_{1}}.}
\end{cases}\label{eq:e_h}
\end{equation}

To sum up, the dynamics of the UAV's battery level can be represented
as 
\begin{equation}
E(t+1)=\textrm{min}\left(E(t)+e_{\textrm{h}}(t)-e_{\textrm{c}}(t),E_{\max}\right).\label{eq:e_update}
\end{equation}

\subsection{Problem Formulation}

In this solar-powered UAV-assisted data collection problem, our goal
is to minimize the time-averaged expected total AoI of SNs and energy
consumption of the UAV. To this end, we jointly optimize the UAV's
trajectory, the schedule of SNs, as well as the offloading strategy
to minimize the system cost, and formulate an optimization problem
as follows:\begin{subequations}

\begin{align}
\text{P1: }\min_{\boldsymbol{p},\boldsymbol{b},\boldsymbol{q}}\quad & \frac{1}{T}\mathbb{E}\left[\sum_{t=1}^{T}\left(\omega_{1}\sum_{n=1}^{N}\delta_{n}(t)+\omega_{2}e_{\textrm{c}}(t)\right)\right],\label{objective}\\
\text{s.t.}\quad & \boldsymbol{C}_{u}(1)=\boldsymbol{C}_{0},\label{eq:initial-position}\\
 & 0\leq v_{s}(t)\leq v_{s}^{\textrm{max}},\label{eq:speed}\\
 & 0\leq\phi(t)\leq2\pi,\label{eq:angle-velocity}\\
 & |\triangle\phi(t)|\leq\triangle\phi_{\max},t\geq2,\label{eq:maximum-angle}
\end{align}
\end{subequations}where $\omega_{1}$ and $\omega_{2}$ represent
the weights of the SNs' average AoI and energy consumption, respectively.
Constraint (\ref{eq:initial-position}) specifies that the UAV takes
off from the DC. \textcolor{black}{The speed constraint (\ref{eq:speed})
and direction constraints (\ref{eq:angle-velocity}) and (\ref{eq:maximum-angle})
guarantee that the UAV adheres to the kinematic restrictions.} The
aforementioned stochastic optimization problem is exceedingly difficult
to solve due to the unknown environmental dynamics, including the
stochastic sampling rates of SNs, the random solar energy arrivals
of the UAV, and the uncertain A2G channel. To deal with this problem,
we propose a reinforcement learning-based algorithm that enables the
UAV to autonomously learn from its interactions with the environment,
thereby jointly optimizing its trajectory, the scheduling of SNs,
and the offloading strategy. The following sections will delve into
a detailed discussion of this approach.

\section{Meta-reinforcement learning Approach \label{sec:Meta-RL}}

In this section, we begin by reformulating the data collection problem
in a solar-powered UAV-assisted IoT network as a MDP. Subsequently,
we propose a compound-action deep reinforcement learning (CADRL) algorithm
to address the MDP, which can handle discrete and continuous actions
simultaneously in the UAV's action space. Furthermore, we introduce
the meta-learning method to enhance the generalization of the CADRL-based
algorithm for new tasks.

\subsection{MDP Formulation}

Most often, an MDP is described by $\{\mathcal{S},\mathcal{A},r,\mathcal{P}\}$,
where $\mathcal{S},\mathcal{A},r$, and $\mathcal{P}$ stand for the
state space, action space, reward function, and state transition function,
respectively. Detailed definitions of state, action, reward function,
and state transition are provided below.

\subsubsection{State}

The state in time slot $t$ is denoted as $\boldsymbol{s}(t)=(\boldsymbol{\Psi}(t),v_{s}(t),\phi(t-1),\boldsymbol{Y}(t),\boldsymbol{U}(t),\boldsymbol{\delta}(t),E(t))$,
where
\begin{itemize}
\item $\boldsymbol{\Psi}(t)$ denotes the set of relative positions between
the UAV's ground projection at time slot $t$ and all the nodes, i.e.,
$\boldsymbol{\Psi}(t)=(\boldsymbol{\psi}_{0}(t),\boldsymbol{\psi}_{1}(t),\ldots,\boldsymbol{\psi}_{N}(t))$,
where $\boldsymbol{\psi}_{n}(t)=\boldsymbol{C}_{u}(t)-\boldsymbol{C}_{n}=(x_{u}(t)-x_{n},y_{u}(t)-y_{n}),n\in\mathcal{N}^{+}$.
\item $v_{s}(t)$ denotes the UAV's speed at the start of time slot $t$.
\item $\phi(t-1)$ denotes the UAV's direction of velocity at time slot
$t-1$.
\item $\boldsymbol{Y}(t)$ is the set of lifetime of the update data packet
in time slot $t$ for all SNs, i.e., $\boldsymbol{Y}(t)=(Y_{1}(t),Y_{2}(t),\ldots,Y_{N}(t))$.
\item $\boldsymbol{U}(t)$ is the set of lifetime of the update data packet
at the UAV in time slot $t$ for all SNs, i.e., $\boldsymbol{U}(t)=(U_{1}(t),U_{2}(t),\ldots,U_{N}(t))$.
\item $\boldsymbol{\delta}(t)$ is the set of AoI values in time slot $t$
for all SNs, i.e., $\boldsymbol{\delta}(t)=(\delta_{1}(t),\delta_{2}(t),\ldots,\delta_{N}(t))$.
\item $E(t)$ denotes the UAV's energy level in time slot $t$. 
\end{itemize}

\subsubsection{Action}

The UAV's action in time slot $t$ is characterized by its speed $v_{s}(t+1)$
at the start of time slot $t+1$ and direction $\phi(t)$, the scheduling
of SNs $b(t)$, and the offloading decision $q(t)$, i.e., $\boldsymbol{a}(t)=(v_{s}(t+1),\phi(t),b(t),q(t))$,
where $v_{s}(t+1)$ and $\phi(t)$ are continuous, while $b(t)$ and
$q(t)$ are discrete.

\subsubsection{State Transition}

We detail the transition of each element in $\boldsymbol{s}(t)$.
The update of the relative position of the UAV's projection on the
ground to the node $n$, $\boldsymbol{\psi}_{n}(t)$ $n\in\mathcal{N}^{+}$,
relies on the UAV's location, speed, and direction of the velocity,
which can be expressed as 
\begin{align}
\boldsymbol{\psi}_{n}(t+1)= & \boldsymbol{\psi}_{n}(t)+\frac{v_{s}(t)+v_{s}(t+1)}{2}\tau_{0}(\cos\phi(t),\sin\phi(t)).\label{eq:update_rela_loc}
\end{align}

The lifetime update of the update data packet at each SN depends
on the arrival status of the packet. In particular, if a new packet
is received at SN $n$, its lifetime at the SN is initialized to zero;
otherwise, the lifetime at the SN is increased by one.  The update
of the lifetime of the update data packet at the SN is given in Eq.
(\ref{eq:SN_life-1}).

The lifetime update of the update data packet for each SN at the UAV
depends on the scheduling status $z_{n}(t)$ from SN $n\in\mathcal{N}$
to the UAV. Specifically, if the update status of SN $n$ is successfully
transmitted to the UAV, the lifetime of SN $n$ at the UAV is set
to its lifetime at SN $n$ plus one; otherwise, the lifetime at the
UAV is increased by one. The update of the lifetime of the update
data packet at the UAV is given in Eq. (\ref{eq:UAV_life1}).

The AoI update for each SN depends on the offloading status $o(t)$
from the UAV to the DC. Due to the channel's stochastic characteristics,
the scheduled SN's AoI may not decrease. Specifically, if the UAV
successfully offloads the packet of SN $n$ in its buffer to the DC
within time slot $t$, the AoI of SN $n$ is updated to the lifetime
of SN $n$ at the UAV; otherwise, the AoI of SN $n$ is incremented
by one. The AoI update is presented in Eq. (\ref{eq:AoI}).

The update of the UAV's energy level is dependent on the energy consumption
and energy arrival status, which is represented in Eq. (\ref{eq:e_update}).

\subsubsection{Reward}

The UAV's reward in time slot $t$ is defined to minimize the weighted
sum of the average AoI of SNs and the energy consumption of the UAV,
and it is expressed as $r(t)=-\frac{1}{T}(\omega_{1}\sum_{n=1}^{N}\delta_{n}(t)+\omega_{2}e_{\textrm{c}}(t))$.

The goal of the MDP is to discover an optimal policy that, beginning
with the initial state $s(1)$, maximizes the expected total reward
during $T$ time slots. In particular, the optimal policy is the solution
of the MDP, i.e.,
\begin{equation}
\pi^{*}=\arg\max_{\pi}\mathbb{E}\left[\sum_{t=1}^{T}r(t)|s(1)\right],\label{eq:optimal_policy}
\end{equation}
where the expectation is relative to the distribution of the sequence
of states and actions following the policy $\pi$ and the transition
probabilities. 

\subsection{Meta-learning-based Compound-action DRL Approach}

\subsubsection{Compound-action DRL Approach}

\begin{figure}[tb]
\includegraphics[width=0.5\textwidth]{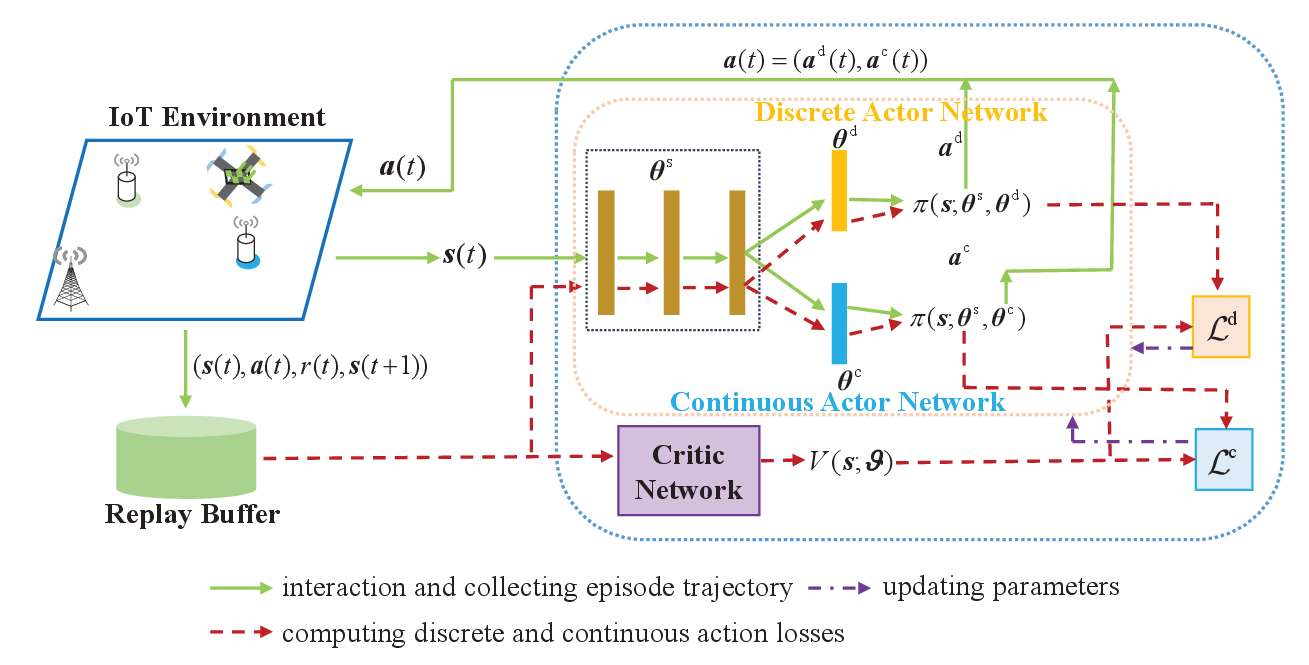}\centering\caption{The architecture of the CADRL approach.\label{fig:Hybrid-action-DRL-architecture}}
\end{figure}

In this subsection, we first introduce the architecture of the CADRL-based
algorithm shown in Fig. \ref{fig:Hybrid-action-DRL-architecture},
which can manage both discrete and continuous actions. This architecture
is based on the actor-critic architecture but contains two parallel
actor networks. These two actor networks are utilized to perform discrete-action
selection and continuous-action selection, respectively. In particular,
the two actor networks share the first few layers with parameters
$\boldsymbol{\theta}^{\textrm{s}}$ to extract the features from the
input state $\boldsymbol{s}$. Then, the single-stream neural network
is partitioned into two streams, forming the discrete actor network's
output layer with parameters $\boldsymbol{\theta}^{\textrm{d}}$ and
the continuous actor network's output layer with parameters $\boldsymbol{\theta}^{\textrm{c}}$,
respectively. A stochastic policy $\pi(\boldsymbol{s};\boldsymbol{\theta}^{\textrm{s}},\boldsymbol{\theta}^{\textrm{d}})$
is learned by the discrete actor network for selecting discrete actions
$\boldsymbol{a}^{\textrm{d}}$, and the continuous actor network learns
a stochastic policy $\pi(\boldsymbol{s};\boldsymbol{\theta}^{\textrm{s}},\boldsymbol{\theta}^{\textrm{c}})$
for determining continuous actions $\boldsymbol{a}^{\textrm{c}}$.

To generate the stochastic policy $\pi(\boldsymbol{s};\boldsymbol{\theta}^{\textrm{s}},\boldsymbol{\theta}^{\textrm{d}})$,
the discrete actor network outputs a vector of values $f=(f_{\boldsymbol{a}_{1}^{\textrm{d}}},f_{\boldsymbol{a}_{2}^{\textrm{d}}},\ldots,f_{\boldsymbol{a}_{A_{\textrm{d}}}^{\textrm{d}}})$
for the $A_{\textrm{d}}$ discrete actions, where $A_{\textrm{d}}=2(N+1)$
is the dimension of the discrete action space. Then, a discrete action
$\boldsymbol{a}^{\textrm{d}}$ is randomly sampled from the distribution
represented by $\textrm{softmax}(f)$. The continuous actor network
outputs both the mean and variance of the Gaussian distribution to
form the stochastic policy $\pi(\boldsymbol{s};\boldsymbol{\theta}^{\textrm{s}},\boldsymbol{\theta}^{\textrm{c}})$
for continuous actions $\boldsymbol{a}^{\textrm{c}}$. Both the discrete
and continuous actor networks are updated using the trust region policy
optimization (TRPO) method in this study. The stochastic policy $\pi(\boldsymbol{s};\boldsymbol{\theta}^{\textrm{s}},\boldsymbol{\theta}^{\textrm{d}})$
and the stochastic policy $\pi(\boldsymbol{s};\boldsymbol{\theta}^{\textrm{s}},\boldsymbol{\theta}^{\textrm{c}})$
are updated by minimizing their loss functions, respectively. The
loss function of the discrete policy $\pi(\boldsymbol{s};\boldsymbol{\theta}^{\textrm{s}},\boldsymbol{\theta}^{\textrm{d}})$
is given by \cite{schulman2015TRPO}
\begin{align}
\mathcal{L}^{\textrm{d}}(\boldsymbol{\theta}^{\textrm{s}},\boldsymbol{\theta}^{\textrm{d}})= & \mathbb{E}\left[\frac{\pi(\boldsymbol{a}^{\textrm{d}}(t)|\boldsymbol{s}(t);\boldsymbol{\theta}^{\textrm{s}},\boldsymbol{\theta}^{\textrm{d}})}{\pi(\boldsymbol{a}^{\textrm{d}}(t)|\boldsymbol{s}(t);\boldsymbol{\theta}_{\textrm{old}}^{\textrm{s}},\boldsymbol{\theta}_{\textrm{old}}^{\textrm{d}})}\hat{A}(\boldsymbol{s}(t))\right],\nonumber \\
s.t.\quad & \mathbb{E}\left[\textrm{KL}[\pi(\cdot|\boldsymbol{s}(t);\boldsymbol{\theta}_{\textrm{old}}^{\textrm{s}},\boldsymbol{\theta}_{\textrm{old}}^{\textrm{d}}),\pi(\cdot|\boldsymbol{s}(t);\boldsymbol{\theta}^{\textrm{s}},\boldsymbol{\theta}^{\textrm{d}})]\right]\leq\epsilon,\label{eq:loss_discrete_actor}
\end{align}
where $\mathbb{E}[\cdot]$ denotes the expectation for a limited batch
of experiences, $(\boldsymbol{\theta}_{\textrm{old}}^{\textrm{s}},\boldsymbol{\theta}_{\textrm{old}}^{\textrm{d}})$
is the parameters before the update, $\textrm{KL}[\pi_{1},\pi_{2}]$
represents the KL divergence between two policies $\pi_{1}$ and $\pi_{2}$,
$\epsilon$ denotes the maximum KL divergence constraint. Similarly,
the loss function of the continuous policy $\pi(\boldsymbol{s};\boldsymbol{\theta}^{\textrm{s}},\boldsymbol{\theta}^{\textrm{c}})$
is given by \cite{schulman2015TRPO}
\begin{align}
\mathcal{L}^{\textrm{c}}(\boldsymbol{\theta}^{\textrm{s}},\boldsymbol{\theta}^{\textrm{c}})= & \mathbb{E}\left[\frac{\pi(\boldsymbol{a}^{\textrm{c}}(t)|\boldsymbol{s}(t);\boldsymbol{\theta}^{\textrm{s}},\boldsymbol{\theta}^{\textrm{c}})}{\pi(\boldsymbol{a}^{\textrm{c}}(t)|\boldsymbol{s}(t);\boldsymbol{\theta}_{\textrm{old}}^{\textrm{s}},\boldsymbol{\theta}_{\textrm{old}}^{\textrm{c}})}\hat{A}(\boldsymbol{s}(t))\right],\nonumber \\
s.t.\quad & \mathbb{E}\left[\textrm{KL}[\pi(\cdot|\boldsymbol{s}(t);\boldsymbol{\theta}_{\textrm{old}}^{\textrm{s}},\boldsymbol{\theta}_{\textrm{old}}^{\textrm{c}}),\pi(\cdot|\boldsymbol{s}(t);\boldsymbol{\theta}^{\textrm{s}},\boldsymbol{\theta}^{\textrm{c}})]\right]\leq\epsilon.\label{eq:loss_continu_actor}
\end{align}

In the CADRL architecture, there is one critic network with parameters
$\boldsymbol{\vartheta}$ that is utilized to estimate the state-value
function $V(\boldsymbol{s};\boldsymbol{\vartheta})$. The critic network
undergoes updates by minimizing the loss function, which is defined
as
\begin{equation}
\mathcal{L}(\boldsymbol{\vartheta})=(V^{\textrm{target}}(t)-V(\boldsymbol{s}(t);\boldsymbol{\vartheta}))^{2},\label{eq:critic-loss}
\end{equation}
where $V^{\textrm{target}}(t)=\hat{A}(t)+V(s(t))$, and $\hat{A}(t)$
is the generalized advantage estimation (GAE) that is expressed by
\cite{schulman2015-GAE}
\begin{equation}
\hat{A}(t)=r(t)+\gamma V(\boldsymbol{s}(t+1))-V(\boldsymbol{s}(t))+\gamma\lambda\hat{A}(t-1),\label{eq:GAE}
\end{equation}
where $\lambda$ is the GAE parameter, $\gamma$ represents the discount
factor, and $\hat{A}(0)=0$.

Algorithm \ref{alg:Hybrid-action-TRPO} outlines the process of the
CADRL-based solar-powered UAV-assisted data collection algorithm.
The process begins with the initialization of the maximum KL divergence
$\epsilon$, replay buffer $D$, as well as the parameters of the
actor network $(\boldsymbol{\theta}^{\textrm{s}},\boldsymbol{\theta}^{\textrm{d}},\boldsymbol{\theta}^{\textrm{c}})$
and critic network $\boldsymbol{\vartheta}$ (Line 1). In time slot
$t$, the UAV selects the discrete action component $\boldsymbol{a}^{\textrm{d}}(t)\sim\pi(\boldsymbol{s}(t);\boldsymbol{\theta}^{\textrm{s}},\boldsymbol{\theta}^{\textrm{d}})$
and the continuous action component $\boldsymbol{a}^{\textrm{c}}(t)\sim\pi(\boldsymbol{s}(t);\boldsymbol{\theta}^{\textrm{s}},\boldsymbol{\theta}^{\textrm{c}})$
according to the current state $\boldsymbol{s}(t)$, conducts the
action $\boldsymbol{a}(t)=(\boldsymbol{a}^{\textrm{d}}(t),\boldsymbol{a}^{\textrm{c}}(t))$,
receives a reward $r(t)$, and transitions a new state $\boldsymbol{s}(t+1)$.
Then, the UAV places the experience ($\boldsymbol{s}(t),\boldsymbol{a}(t),r(t)$,$\boldsymbol{s}(t+1)$)
in the replay buffer (Lines 5\textasciitilde 8). To update the actor
and critic networks, a mini-batch of experiences is sampled from the
replay buffer. The advantage function is computed as (\ref{eq:GAE}).
The loss functions of the discrete and continuous actor networks are
calculated according to (\ref{eq:loss_discrete_actor}) and (\ref{eq:loss_continu_actor}),
respectively, and are minimized to update the two actor networks.
The critic network undergoes updates through the minimization of its
loss function (\ref{eq:critic-loss}) (Lines 9\textasciitilde 12). 

\begin{algorithm}[tb]
\caption{CADRL-based solar-powered UAV-assisted data collection algorithm.
\label{alg:Hybrid-action-TRPO}}

\begin{algorithmic}[1]

\STATE Initialize the maximum KL divergence $\epsilon$, the replay
buffer $D$, the discrete and continuous actor networks parameters
$(\boldsymbol{\theta}^{\textrm{s}},\boldsymbol{\theta}^{\textrm{d}},\boldsymbol{\theta}^{\textrm{c}})$,
and the critic network parameters $\boldsymbol{\vartheta}$;

\FOR{ $\textrm{ep}=1:E$ }

\STATE Initialize the environment;

\FOR{$t=1:T$}

\STATE Observe the state $\boldsymbol{s}(t)$;

\STATE Select a discrete action $\boldsymbol{a}^{\textrm{d}}(t)\sim\pi(\boldsymbol{s}(t);\boldsymbol{\theta}^{\textrm{s}},\boldsymbol{\theta}^{\textrm{d}})$
and a continuous action $\boldsymbol{a}^{\textrm{c}}(t)\sim\pi(\boldsymbol{s}(t);\boldsymbol{\theta}^{\textrm{s}},\boldsymbol{\theta}^{\textrm{c}})$;

\STATE Conduct action $\boldsymbol{a}(t)=(\boldsymbol{a}^{\textrm{d}}(t),\boldsymbol{a}^{\textrm{c}}(t))$,
receive the reward $r(t)$, and transition to the next state $\boldsymbol{s}(t+1)$;

\STATE Store experience $(\boldsymbol{s}(t),\boldsymbol{a}(t),r(t),\boldsymbol{s}(t+1))$
in $D$;

\STATE Get out a mini-batch of $J$ experiences $(\boldsymbol{s}(j),\boldsymbol{a}(j),r(j),\boldsymbol{s}(j+1))$
from $D$ to update the discrete and continuous actor networks;

\STATE $\pi(\boldsymbol{s};\boldsymbol{\theta}_{\textrm{old}}^{\textrm{s}},\boldsymbol{\theta}_{\textrm{old}}^{\textrm{d}})=\pi(\boldsymbol{s};\boldsymbol{\theta}^{\textrm{s}},\boldsymbol{\theta}^{\textrm{d}})$
and $\pi(\boldsymbol{s};\boldsymbol{\theta}_{\textrm{old}}^{\textrm{s}},\boldsymbol{\theta}_{\textrm{old}}^{\textrm{c}})=\pi(\boldsymbol{s};\boldsymbol{\theta}^{\textrm{s}},\boldsymbol{\theta}^{\textrm{c}})$;

\STATE Compute the advantage function depends on (\ref{eq:GAE});

\STATE The discrete and continuous actor networks are updated by
minimizing the loss functions (\ref{eq:loss_discrete_actor}) and
(\ref{eq:loss_continu_actor}), respectively, and the critic network
undergoes updates through the minimization of its loss function (\ref{eq:critic-loss}).

\ENDFOR

\ENDFOR

\end{algorithmic}
\end{algorithm}

\subsubsection{Meta-learning-based Compound-action DRL}

The CADRL-based algorithm is applicable to a specific task where
the number and positions of SNs are fixed. When the task changes,
the policy learned by the CADRL-based algorithm is not applicable
to the new task. The use of a meta-learning-based DRL method brings
the advantage of quick adaptation to new tasks, drawing on previous
knowledge from prior experiences, and minimising the need for extensive
training data \cite{thrun1998learning}. In parallel, model-agnostic
meta-learning (MAML) is a gradient-based meta-RL approach that concentrates
on parameter optimization of the meta-policy during meta-training,
serving as a solid starting point for unforeseen tasks \cite{finn2017model}.
Hence, we utilize MAML to enhance the generalization of the DRL method
across different tasks and propose a meta-learning-based compound-action
deep reinforcement learning (MLCADRL) algorithm. 

\begin{figure*}[tb]
\includegraphics[width=0.8\textwidth]{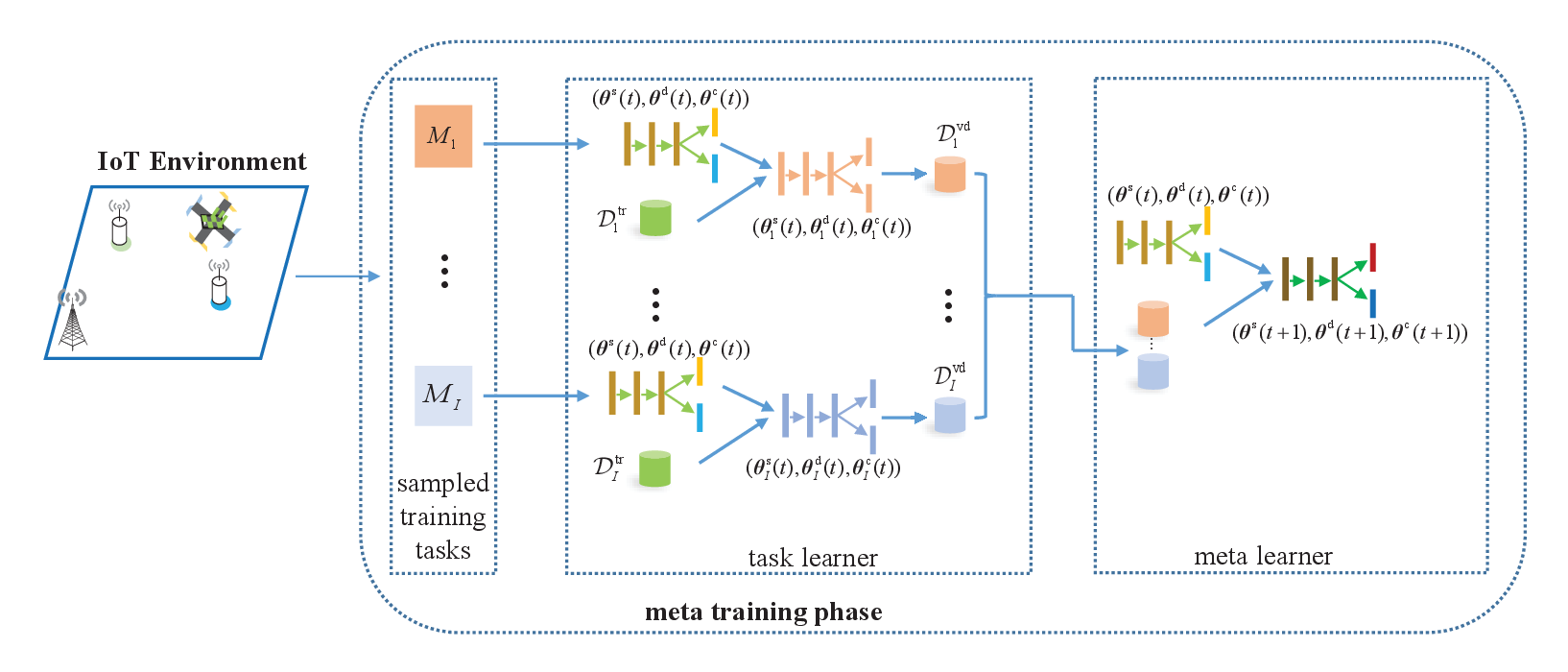}\centering

\caption{The training phase of meta-learning for hybrid-action DRL.\label{fig:meta-training-phase}}
\end{figure*}

An agent in MAML endeavors to acquire a meta-policy with parameters
$\boldsymbol{\theta}=(\boldsymbol{\theta}^{\textrm{s}},\boldsymbol{\theta}^{\textrm{d}},\boldsymbol{\theta}^{\textrm{c}})$
across a multitude of tasks from a given task distribution $p(M)$.
Each task $M_{i}\sim p(M)$ can be described as an MDP model defined
by $(\mathcal{S}_{i},\mathcal{A}_{i},r_{i},\mathcal{P}_{i})$, where
$\mathcal{S}_{i}$, $\mathcal{A}_{i}$, $r_{i}$, and $\mathcal{P}_{i}$
denote the state space, action space, reward function, and state transition
function of task $M_{i}$, respectively. As depicted in Fig. \ref{fig:meta-training-phase},
the training process for the MLCADRL-based algorithm consists of two
alternating learning phases: the inner-loop task learner and the outer-loop
meta learner. The meta parameters $\boldsymbol{\theta}$ in both loops
are optimized by using gradient descent. In the inner-loop, the task
learner is initialized with the meta parameters $\boldsymbol{\theta}$.
It computes the updated parameters $\boldsymbol{\theta}_{i}$ of the
task learner for each training task $M_{i}$ using the training data
$\mathcal{D}_{i}^{\textrm{tr}}$, which is collected by the meta-policy
$\boldsymbol{\theta}$, which is expressed as
\begin{equation}
\boldsymbol{\theta}_{i}\leftarrow\boldsymbol{\theta}-\alpha_{1}\nabla_{\boldsymbol{\theta}}\mathcal{L}(\boldsymbol{\theta},\mathcal{D}_{i}^{\textrm{tr}}),\label{eq:inner_updata}
\end{equation}
where $\alpha_{1}$ is the task learner's learning rate and $\mathcal{L}$
denotes the loss function of an RL learning method. Then, the task
learner utilizes validation data $\mathcal{D}_{i}^{\textrm{vd}}$
sampled from trajectories collected with updated parameters $\boldsymbol{\theta}_{i}$
for task $M_{i}$ to estimate the loss function, expressed as
\begin{equation}
\mathcal{L}_{M_{i}}(\boldsymbol{\theta}_{i},\mathcal{D}_{i}^{\textrm{vd}})=\mathcal{L}_{M_{i}}(\boldsymbol{\theta}-\alpha_{1}\nabla_{\boldsymbol{\theta}}\mathcal{L}(\boldsymbol{\theta},\mathcal{D}_{i}^{\textrm{tr}}),\mathcal{D}_{i}^{\textrm{vd}}).\label{eq:loss function}
\end{equation}
In the outer-loop, as the policy is updated for each training task,
the meta learner accumulates the loss $\mathcal{L}_{M_{i}}(\boldsymbol{\theta}_{i},\mathcal{D}_{i}^{\textrm{vd}})$
and carries out a meta-gradient update on the meta parameters $\boldsymbol{\theta}$
as
\begin{equation}
\boldsymbol{\theta}\leftarrow\boldsymbol{\theta}-\alpha_{2}\nabla_{\boldsymbol{\theta}}\sum_{M_{i}\sim p(M)}\mathcal{L}_{M_{i}}(\boldsymbol{\theta}_{i},\mathcal{D}_{i}^{\textrm{vd}}),\label{eq:outer_update}
\end{equation}
where $\alpha_{2}$ is the learning rate of the meta learner. These
processes iterate for $L_{\textrm{meta}}$ times. Upon the completion
of meta-training, the meta-policy can be employed as the initial policy
for new tasks, facilitating rapid adaptation to new tasks.

The solar-powered UAV-assisted data collection algorithm, based on
MLCADRL, is detailed in Algorithm \ref{alg:Algorithm2}. First, the
task distribution $p(M)$, the number of meta-iterations $L_{\textrm{meta}}$,
the number of tasks of each meta-iteration $I$, the number of trajectories
sampled for each task $K$, and the parameters of the meta-policy
$\pi(\boldsymbol{\theta}^{\textrm{s}},\boldsymbol{\theta}^{\textrm{d}},\boldsymbol{\theta}^{\textrm{c}})$
are initialized (Line 1). In the meta-training phase, the inner-loop
task learner and outer-loop meta learner are performed alternately.
In particular, in the inner-loop task learner phase, the task-specific
policy is updated by using the vanilla policy gradient algorithm (REINFORCE)
\cite{williams1992simple}. We sample $I$ tasks from the task distribution
(Line 3). The loss functions of the discrete and continuous actor
networks for each task $M_{i}$ are given by
\begin{align}
\mathcal{L}_{i}^{\textrm{i,d}}(\boldsymbol{\theta}^{\textrm{s}},\boldsymbol{\theta}^{\textrm{d}})= & \frac{1}{KT}\sum_{k=1}^{K}\sum_{t=1}^{T}\mathbb{E}\Bigg[\log\pi(\boldsymbol{a}_{i,k}^{\textrm{d}}(t)|\boldsymbol{s}_{i,k}(t);\boldsymbol{\theta}^{\textrm{s}},\boldsymbol{\theta}^{\textrm{d}})\times\nonumber \\
 & \quad\quad\quad\quad\quad\quad\hat{A}(\boldsymbol{s}_{i,k}(t))\Bigg],\label{eq:loss_d_inner}
\end{align}
\\
\begin{align}
\mathcal{L}_{i}^{\textrm{i,c}}(\boldsymbol{\theta}^{\textrm{s}},\boldsymbol{\theta}^{\textrm{c}})= & \frac{1}{KT}\sum_{k=1}^{K}\sum_{t=1}^{T}\mathbb{E}\Bigg[\log\pi(\boldsymbol{a}_{i,k}^{\textrm{c}}(t)|\boldsymbol{s}_{i,k}(t);\boldsymbol{\theta}^{\textrm{s}},\boldsymbol{\theta}^{\textrm{c}})\times\nonumber \\
 & \quad\quad\quad\quad\quad\quad\hat{A}(\boldsymbol{s}_{i,k}(t))\Bigg],\label{eq:loss_c_inner}
\end{align}
where $\boldsymbol{s}_{i,k}(t)$, $\boldsymbol{a}_{i,k}^{\textrm{c}}(t)$,
and $\boldsymbol{a}_{i,k}^{\textrm{d}}(t)$ respectively denote the
state, continuous action, and discrete action of task $M_{i}$ at
time slot $t$ along trajectory $k$. We generate $K$ trajectories
$\mathcal{D}_{i}^{\textrm{tr}}$ by executing the meta-policy on task
$M_{i}$ to calculate the gradients of the loss functions in (\ref{eq:loss_d_inner})
and (\ref{eq:loss_c_inner}). The task-specific policy parameters
$(\boldsymbol{\theta}_{i}^{\textrm{s}},\boldsymbol{\theta}_{i}^{\textrm{d}},\boldsymbol{\theta}_{i}^{\textrm{c}})$
are obtained by performing one or more gradient updates in (\ref{eq:inner_updata})
(Lines 5\textasciitilde 7). In the outer-loop meta-learner phase,
the loss functions of the discrete and continuous actor networks are
given by
\begin{align}
\mathcal{L}^{\textrm{o,d}}(\boldsymbol{\theta}^{\textrm{s}},\boldsymbol{\theta}^{\textrm{d}})= & \sum_{i=1}^{I}\mathcal{L}^{\textrm{d}}(\boldsymbol{\theta}_{i}^{\textrm{s}},\boldsymbol{\theta}_{i}^{\textrm{d}}),\label{eq:loss_d_outer}\\
\mathcal{L}^{\textrm{o,c}}(\boldsymbol{\theta}^{\textrm{s}},\boldsymbol{\theta}^{\textrm{c}})= & \sum_{i=1}^{I}\mathcal{L}^{\textrm{c}}(\boldsymbol{\theta}_{i}^{\textrm{s}},\boldsymbol{\theta}_{i}^{\textrm{d}}).\label{eq:loss_c_outer}
\end{align}
To calculate the gradients of the loss functions in (\ref{eq:loss_d_outer})
and (\ref{eq:loss_c_outer}), the meta-learner summarizes the trajectories
$\mathcal{D}_{i}^{\textrm{vd}}$ sampled using the policy $\pi(\boldsymbol{\theta}_{i}^{\textrm{s}},\boldsymbol{\theta}_{i}^{\textrm{d}},\boldsymbol{\theta}_{i}^{\textrm{c}})$
for each task. The meta-policy is updated by minimizing these loss
functions using the TRPO method (Lines 8 and 10).

\begin{algorithm}[tb]
\caption{Algorithm of the MLCADRL-based for solar-powered UAV-assisted data
collection. \label{alg:Algorithm2}}

\begin{algorithmic}[1]

\STATE Initialize the task distribution $p(M)$, the number of meta-iterations
$L_{\textrm{meta}}$, the number of tasks of each meta-iteration $I$,
the number of trajectories sampled for each task $K$, the parameters
of the compound-action meta-policy $\pi(\boldsymbol{\theta}^{\textrm{s}},\boldsymbol{\theta}^{\textrm{d}},\boldsymbol{\theta}^{\textrm{c}})$;

\renewcommand{\algorithmicrequire}{\textbf{}}

\FOR{ $l_{\textrm{meta}}=1:L_{\textrm{meta}}$ }

\REQUIRE\textbf{\textit{Task learner in the inner-loop}}

\STATE Sample $I$ tasks $M_{i}\sim p(M)$;

\FOR{ each task $M_{i}$ }

\STATE Sample $K$ trajectories $\mathcal{D}_{i}^{\textrm{tr}}$
using $\pi(\boldsymbol{\theta}^{\textrm{s}},\boldsymbol{\theta}^{\textrm{d}},\boldsymbol{\theta}^{\textrm{c}})$;

\STATE Compute inner-loss functions of discrete and continuous actor
networks in (\ref{eq:loss_d_inner}) and (\ref{eq:loss_c_inner})
using $\mathcal{D}_{i}^{\textrm{tr}}$;

\STATE Update task-specific parameters $(\boldsymbol{\theta}_{i}^{\textrm{s}},\boldsymbol{\theta}_{i}^{\textrm{d}},\boldsymbol{\theta}_{i}^{\textrm{c}})$
as in (\ref{eq:inner_updata});

\STATE Collect trajectories $\mathcal{D}_{i}^{\textrm{vd}}$ using
the updated policy $\pi(\boldsymbol{\theta}_{i}^{\textrm{s}},\boldsymbol{\theta}_{i}^{\textrm{d}},\boldsymbol{\theta}_{i}^{\textrm{c}})$
in task $M_{i}$;

\ENDFOR

\textbf{\textit{Meta learner in the outer-loop}}

\STATE Update meta-parameters $(\boldsymbol{\theta}^{\textrm{s}},\boldsymbol{\theta}^{\textrm{d}},\boldsymbol{\theta}^{\textrm{c}})$
by minimizing loss functions of discrete and continuous actor networks
in (\ref{eq:loss_d_outer}) and (\ref{eq:loss_c_outer}) using $\mathcal{D}_{i}^{\textrm{vd}}$;

\ENDFOR

\end{algorithmic}
\end{algorithm}

\section{Simulation Results\label{sec:Simulation-Results}}

In this section, we conduct extensive simulations to gauge the effectiveness
of our CADRL-based and MLCADRL-based algorithms. Firstly, the simulation
setup and baseline algorithms are introduced. We then explore the
CADRL-based algorithm's convergence and performance under different
environmental parameters. Additionally, we delve into the influence
of parameters on the MLCADRL-based algorithm's training and its performance
on new tasks.

\subsection{Simulation Setup}

In our simulation, we randomly distribute SNs within a square area
measuring $200\,\textrm{m}$ on each side. The data DC is situated
at coordinates $(0,160\,\textrm{m})$. Table \ref{tab:System-parameters}
outlines the essential system parameters of the IoT network.
\begin{table}[tb]
\caption{System parameters\label{tab:System-parameters}}

\centering

\begin{tabular}{c|c}
\hline 
Parameter & Value\tabularnewline
\hline 
$P_{\textrm{c}}^{\textrm{t}},v_{s}^{\max},\triangle\phi_{\max}$ & 1 W, 20 m/s, $\frac{\pi}{3}$\tabularnewline
\hline 
$\beta$,$\beta'$,$\beta_{0}$ & $11.95$,$0.14$,$-60\,\textrm{dB}$,\tabularnewline
\hline 
$w$,$\kappa$,$\varsigma$ & $10240$ bits,$0.2$,$2.3$\tabularnewline
\hline 
$\lambda_{0}$,$\sigma^{2}$,$\xi_{\textrm{th}}$ & $0.1$,$-100\,\textrm{dBm}$\cite{abd-elmagidDeepReinforcementLearning2019},$2\,\textrm{dB}$\tabularnewline
\hline 
$H$, $T$, $\tau_{0}$, $\tau_{\textrm{s}}$ & 100 m, 100 slots, 0.5 s, 0.25 s\tabularnewline
\hline 
$P_{s}$,$B$ & $50\,\textrm{mW}$, $5\,\textrm{MHz}$\tabularnewline
\hline 
$n_{r}$,$\chi$,$x_{T}$ & $4$,$0.012$,$0.302$ \cite{RDing_3DTra-Freq-EE-FairC-DRL_2020TWC}\tabularnewline
\hline 
$\rho$,$A$,$x_{s}$ & $1.293\,\textrm{kg/\ensuremath{\textrm{m}^{3}}}$,$0.0314\,\textrm{m}^{2}$,$0.0955$\tabularnewline
\hline 
$d_{0}$,$x_{f}$,$S_{FA}$ & $0.834$,$0.131$,$0.1$\tabularnewline
\hline 
$M$,$g$ & $2\,\textrm{kg}$,$9.8\,\textrm{m/}\textrm{s}^{2}$\tabularnewline
\hline 
$E_{\textrm{th}}^{1}$,$E_{\textrm{th}}^{2}$,$E_{\max}$ & $1e3\,\textrm{J}$,$3e3\,\textrm{J}$,$6e3\,\textrm{J}$\tabularnewline
\hline 
$\eta_{1}$,$S_{u}$,$G_{1}$ & $0.4$,$0.14\,\textrm{m}^{2}$,$1367\,\textrm{W/}\textrm{m}^{2}$
\cite{Fu_Solar-power-uav_TVT2021}\tabularnewline
\hline 
$\varphi_{1}$,$\varphi_{2}$,$h_{1}$ & $0.8978$,$0.2804$,$8000$\tabularnewline
\hline 
$\omega_{1}$,$\omega_{2}$ & $1$,$10$\tabularnewline
\hline 
\end{tabular}
\end{table}

We utilize the NVIDIA GeForce RTX 3080 GPU alongside the Gen Intel(R)
Core(TM) i9-12900K CPU with a clock frequency of 3.20 GHz for simulations.
Our software environment was composed of Torch 1.3.0 and Python 3.6,
running on the Ubuntu 20.04 LTS platform. In the CADRL-based algorithm,
the actor network is composed of three fully connected hidden layers
with $256$ neurons in each layer. Beyond the third hidden layer,
the network splits into two streams: one with the same size as the
discrete action space, representing the discrete stochastic policy,
and the other with the same size as the continuous action dimension,
signifying the continuous stochastic policy. The critic network is
comprised of two fully connected hidden layers, each containing $256$
neurons, and the output layer with one neuron representing the value
function. The input layers of both the actor and critic networks maintain
the same dimensionality as the state in one time slot. In the MLCADRL-based
algorithm, the meta-policy network adheres to the identical architecture
found in the actor network of the CADRL-based algorithm. Further details
of the hyperparameters for both the CADRL-based and MLCADRL-based
algorithms are in Table \ref{tab:Hyperparameters}. 

\begin{table}[tb]
\caption{Hyperparameters of CADRL-based and MLCADRL-based algorithms\label{tab:Hyperparameters}}

\centering

\begin{tabular}{c|c}
\hline 
Parameter & Value\tabularnewline
\hline 
$E$,$D$ & $30000$,$2048$\tabularnewline
\hline 
$J$,$\epsilon$ & $64$,$0.01$\tabularnewline
\hline 
$\lambda$,$\gamma$,$\alpha_{1}$ & $0.95$,$0.99$,$0.1$\tabularnewline
\hline 
$L_{\textrm{meta}}$,$K$,$I$ & $10000$,$5$,$15$\tabularnewline
\hline 
Activation function & Relu\tabularnewline
\hline 
\end{tabular}
\end{table}

The performance of the proposed algorithms is assessed in comparison
to five different baseline algorithms, namely:
\begin{itemize}
\item DQN-based algorithm \cite{DQN}: In this algorithm, we discretize
the continuous actions into discrete actions.
\item DDPG-based algorithm \cite{DDPG}: In this algorithm, we convert the
discrete actions into continuous actions.
\item P-DQN-based algorithm \cite{xiong2018parametrized}: This algorithm
can be seen as a combination of DQN and DDPG to solve the problem
involving continuous and discrete actions. 
\item AoI-based algorithm: In this algorithm, the UAV flies and schedules
a SN during each time slot, guided by the following characteristics:
the SN has updated data packets in its buffer, and the SN possesses
the maximum AoI value at the DC. Once the UAV collects data from the
SN, it promptly forwards the collected information to the DC.
\item TLCADRL-based algorithm \cite{taylor2009transfer}: The algorithm
is a combination of the CADRL-based algorithm and transfer learning.
This approach allows the strategy obtained from similar tasks to be
transferred to new tasks.
\end{itemize}

\subsection{The Performance of CADRL-based Algorithm}

\begin{figure}[tb]
\centering\includegraphics[width=0.45\textwidth]{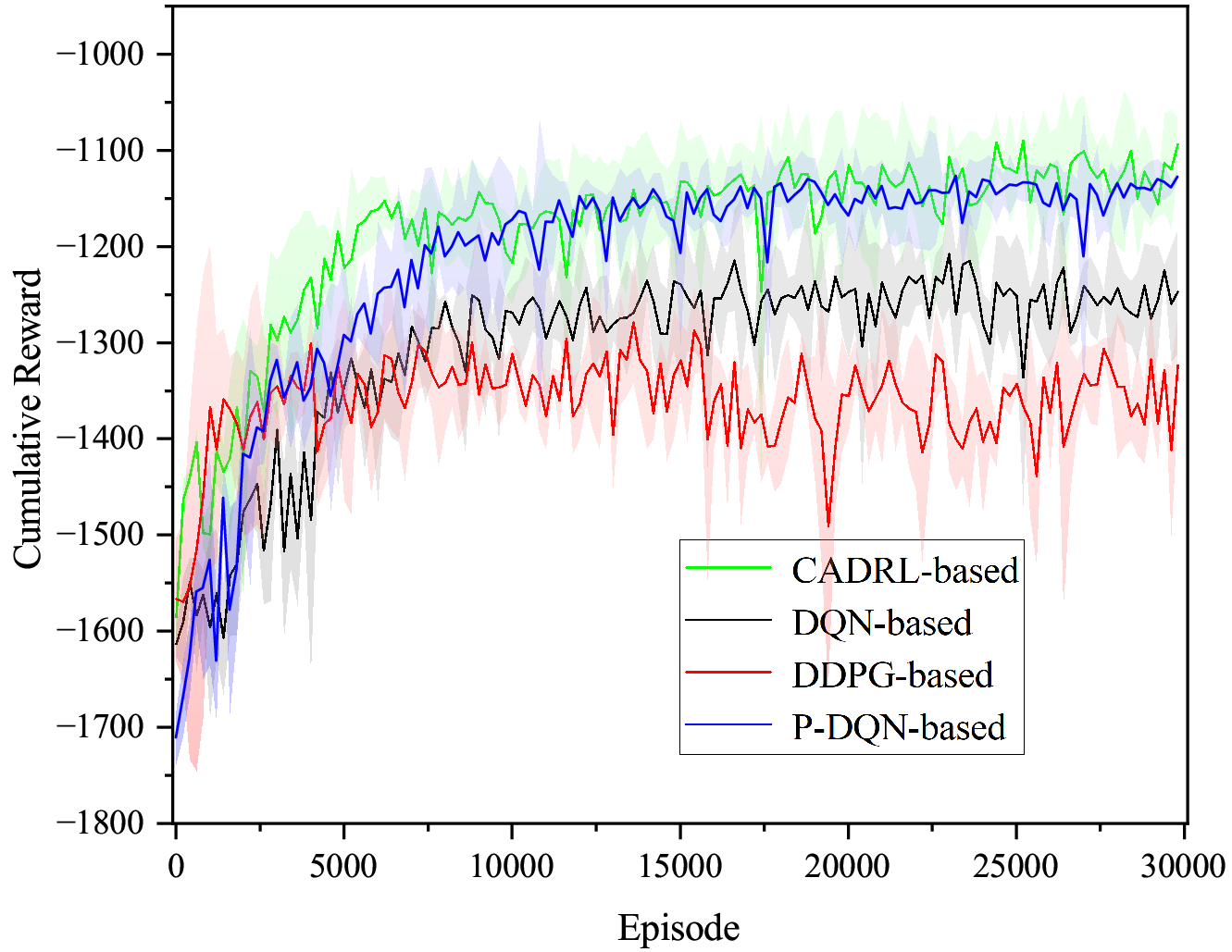}

\caption{Evaluation of convergence performance for the proposed CADRL-based
algorithm and other learning-based algorithms $(\omega_{1}:\omega_{2}=1:10,N=20,\lambda_{0}=0.1\textrm{ and }\lambda_{1}=0.7)$).\label{fig:Evaluation-of-convergence}}
\end{figure}

Fig. \ref{fig:Evaluation-of-convergence} displays the convergence
curves of four algorithms: our proposed CADRL-based algorithm, the
DQN-based algorithm, the DDPG-based algorithm, and the P-DQN-based
algorithm. The horizontal axis is the number of training episodes,
while the vertical axis represents the cumulative reward. It is evident
that our CADRL-based algorithm outperforms the other algorithms. Additionally,
the CADRL-based algorithm's performance matches that of the P-DQN-based
algorithm in the final stages, but with faster convergence. This superiority
can be attributed to its innovative hybrid structure, which enables
this algorithm to effectively address the challenges posed by the
compound action space.

\begin{figure*}[tb]
\subfloat[]{\centering\includegraphics[width=0.33\textwidth]{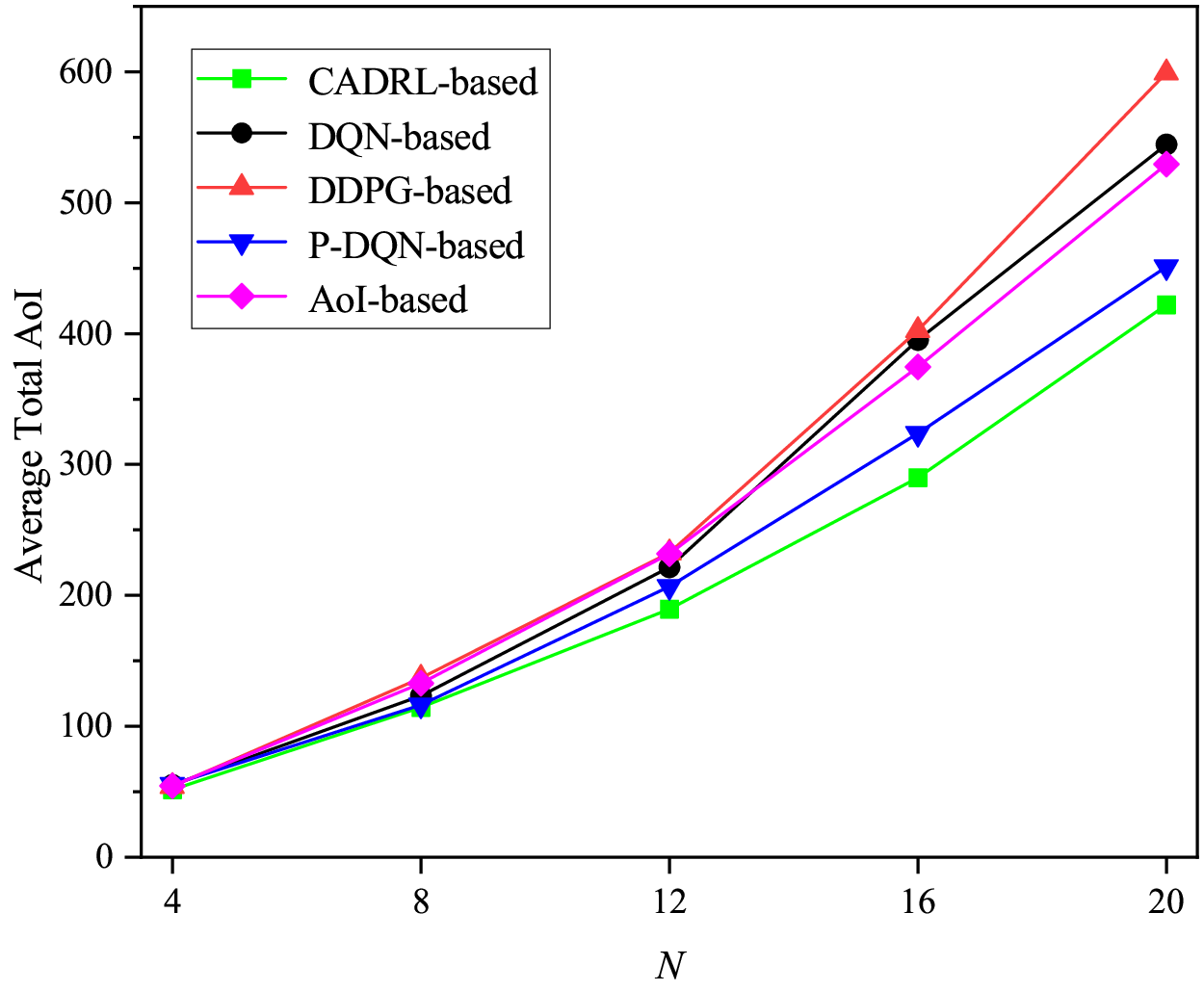}}\subfloat[]{\centering\includegraphics[width=0.33\textwidth]{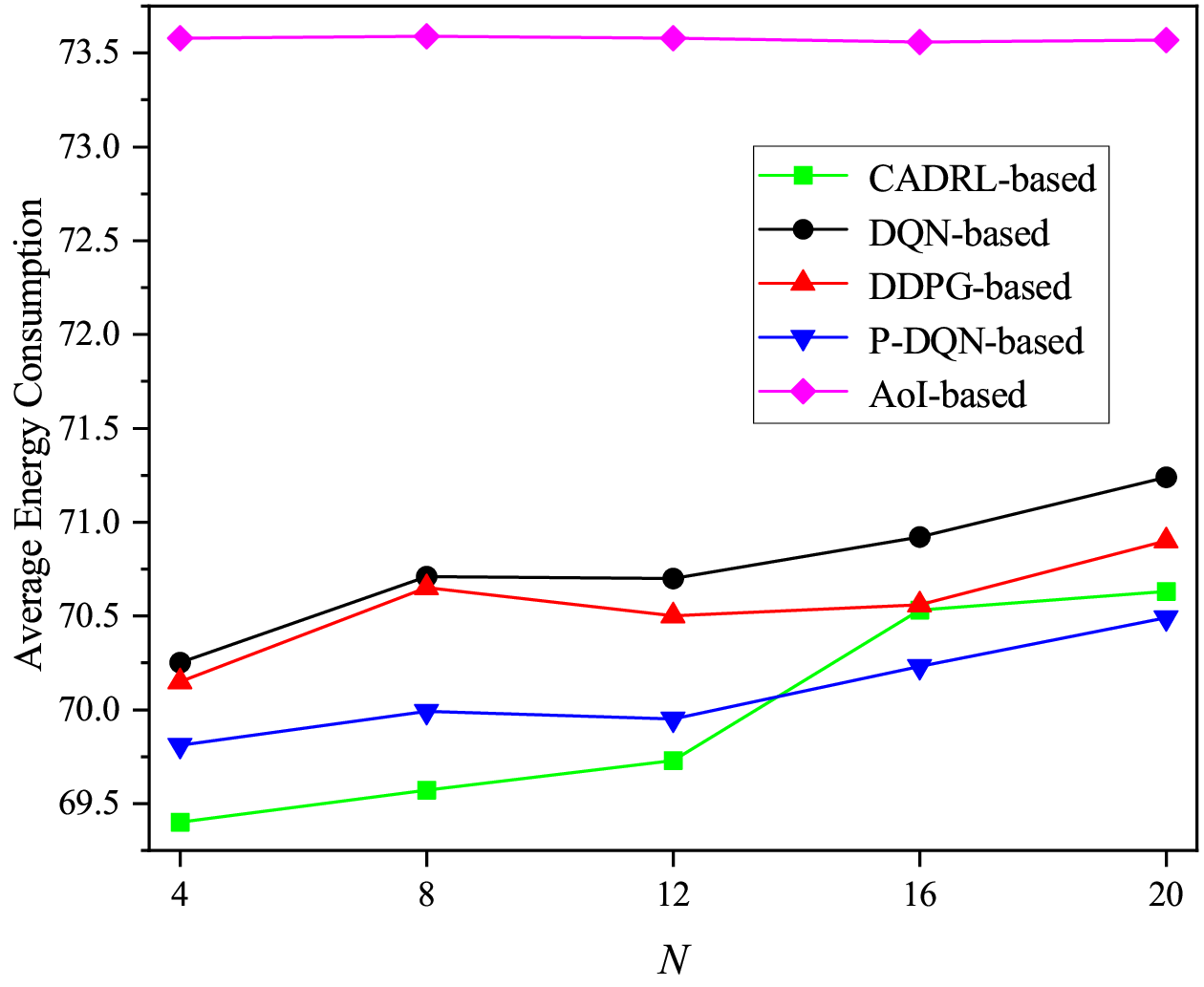}}\subfloat[]{\includegraphics[width=0.33\textwidth]{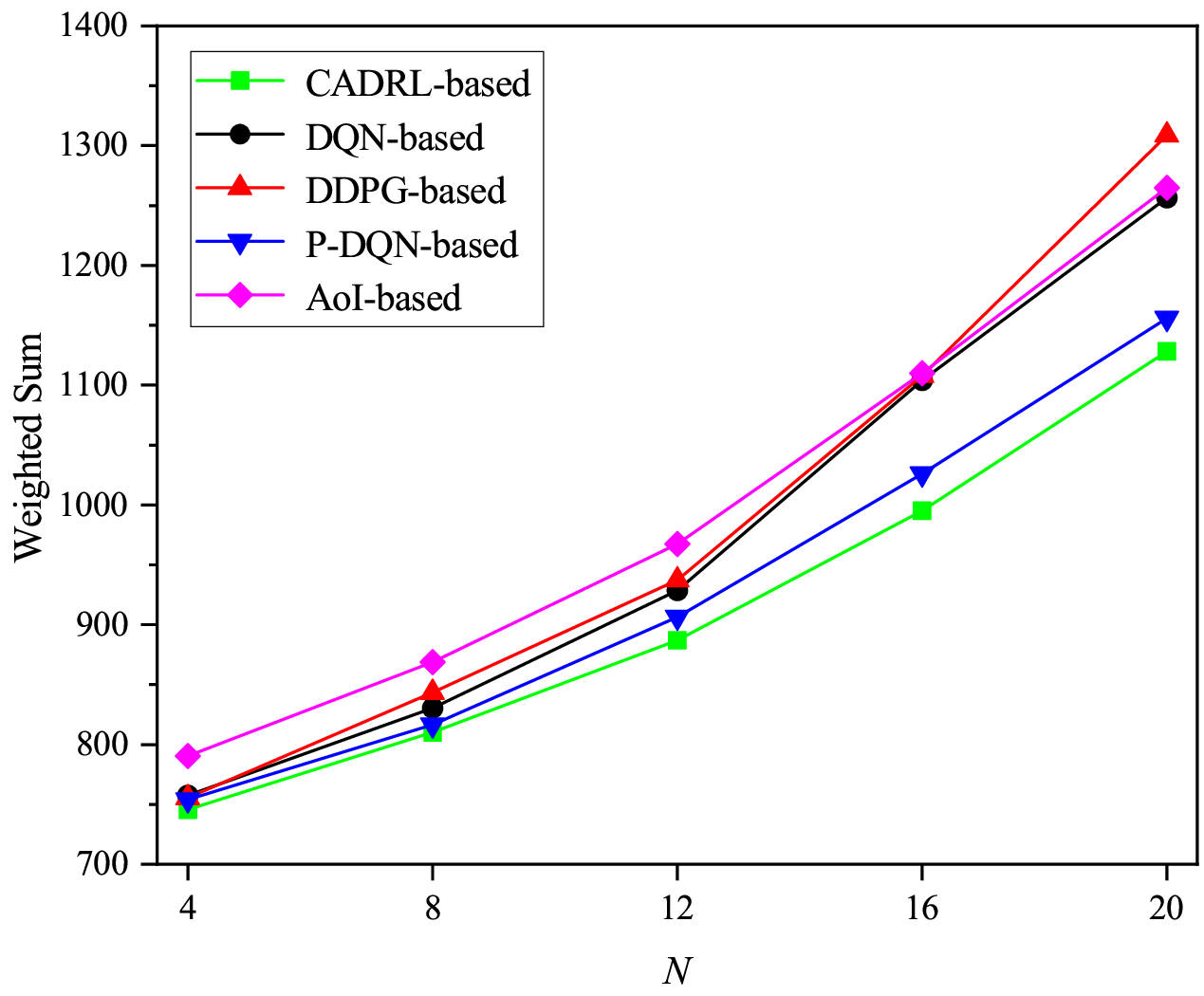}\centering}

\caption{Performance comparison of the proposed CADRL-based algorithm and four
baseline algorithms $(\omega_{1}:\omega_{2}=1:10,\lambda_{0}=0.1\textrm{ and }\lambda_{1}=0.7)$.
(a) the average AoI of SNs in relation to the number of SNs $N$.
(b) The average energy consumption in relation to the number of SNs
$N$. (c) The weighted sum in relation to the number of SNs $N$.
\label{fig:number-SNs}}
\end{figure*}

Fig. \ref{fig:number-SNs}(a) reveals the average AoI of SNs in relation
to the number of SNs $N$. As $N$ grows, a noticeable rise in the
average AoI can be observed. This can be attributed to the constraint
that allows the UAV to schedule only one SN for status updates in
each time slot. Consequently, with a larger $N$, each SN experiences
longer waiting times to update its status, leading to extended AoI
update periods at the DC. Fig. \ref{fig:number-SNs}(b) shows the
UAV's average energy consumption in relation to $N$. As $N$ increases,
the learning-based methods exhibit an increasing trend in average
energy consumption, while the AoI-based algorithm maintains a relatively
constant average energy consumption. Furthermore, the average energy
consumption of the AoI-based algorithm surpasses that of the learning-based
algorithms. This is due to the fact that in the AoI-based algorithm,
the UAV accelerates to its maximum speed with maximum acceleration
to collect the update status from the target SN, resulting in higher
energy consumption as acceleration increases. On the other hand,
the learning-based algorithms effectively control the UAV's velocity,
leading to reduced energy consumption. As $N$ grows, the UAV is required
to adjust its velocity more frequently to adapt to changes in its
flight state. Furthermore, we can see that the weighted sum increases
with $N$ from Fig. \ref{fig:number-SNs}(c), and our proposed CADRL-based
algorithm attains the lowest weighted sum in comparison to the other
algorithms.

\begin{figure*}[tb]
\subfloat[]{\centering\includegraphics[width=0.33\textwidth]{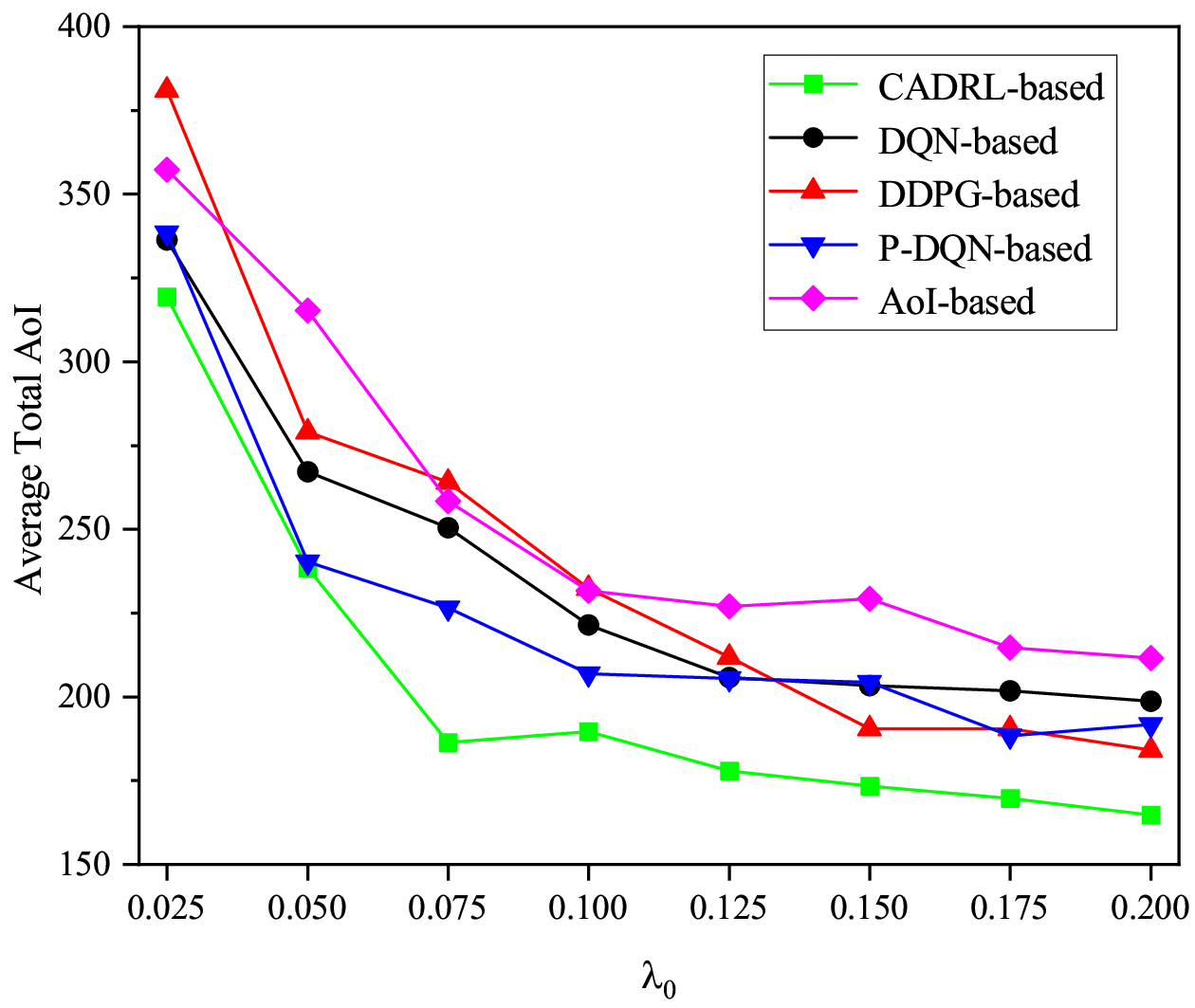}}\subfloat[]{\centering\includegraphics[width=0.33\textwidth]{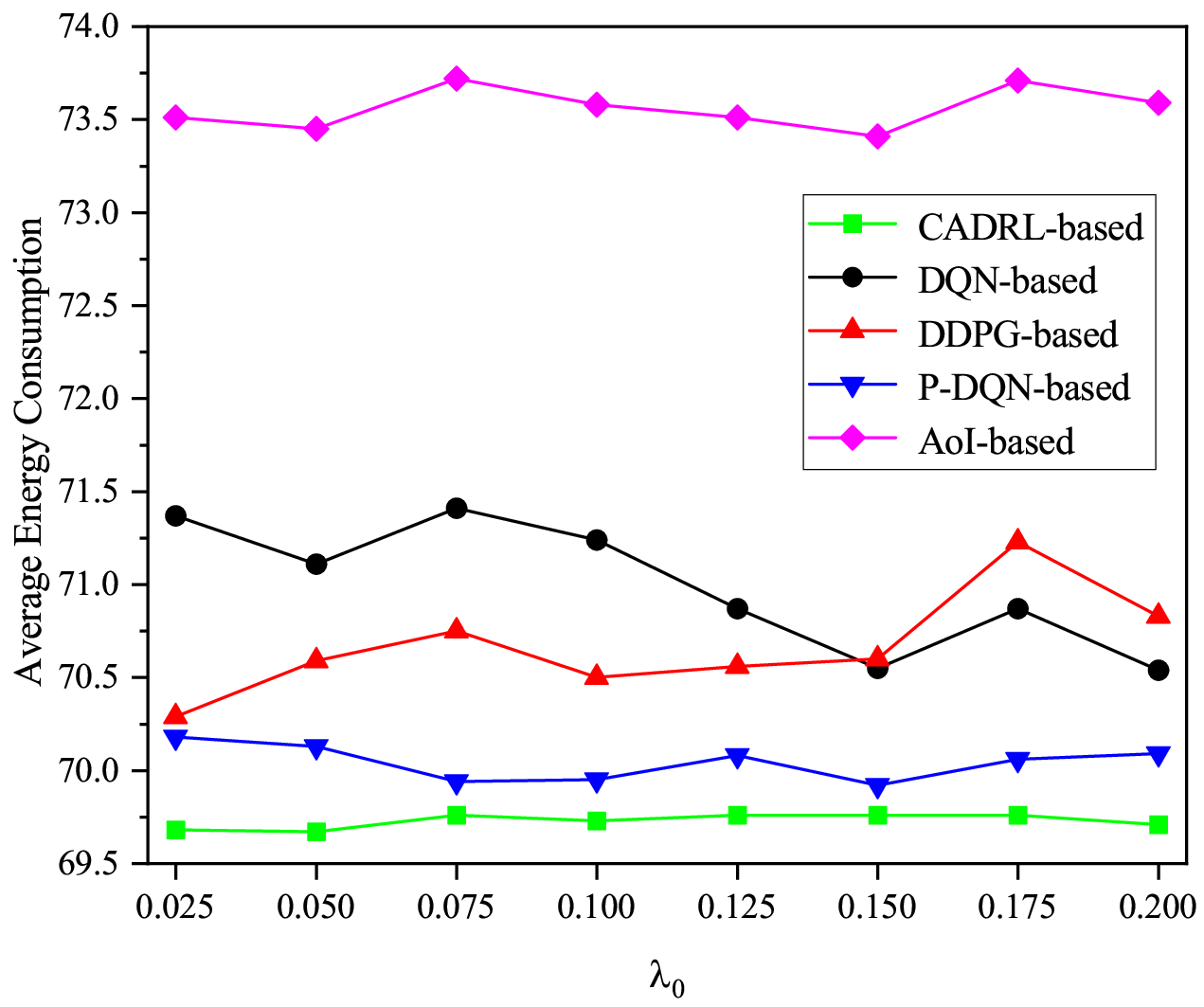}}\subfloat[]{\centering\includegraphics[width=0.33\textwidth]{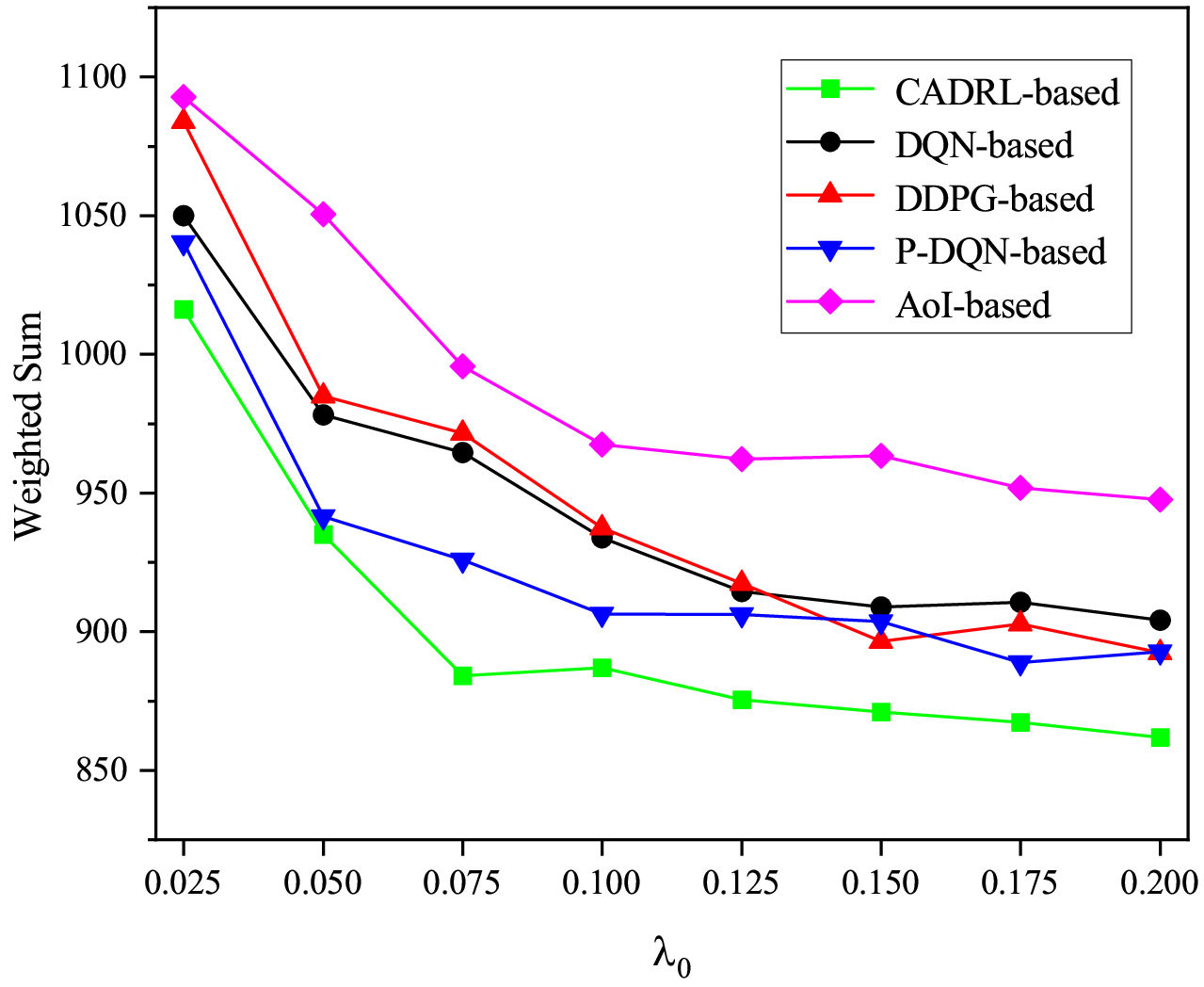}

}\caption{Performance comparison of the proposed CADRL-based algorithm and four
baseline algorithms $(\omega_{1}:\omega_{2}=1:10,N=12\textrm{ and }\lambda_{1}=0.7)$.
(a) the average AoI of SNs in relation to the sample rate $\lambda_{0}$.
(b) The average energy consumption in relation to the sample rate
$\lambda_{0}$. (c) The weighted sum in relation to the sample rate
$\lambda_{0}$. \label{fig:sample-rate}}
\end{figure*}

Fig. \ref{fig:sample-rate}(a) illustrates the relationship between
the SN's sampling rate $\lambda_{0}$ and the average AoI. In Fig.
\ref{fig:sample-rate}(a), it is noticeable that as $\lambda_{0}$
increases, the average AoI of SNs undergoes an initial rapid decrease,
after which the rate of decline gradually slows down, ultimately reaching
a stable state. This is due to the fact that when $\lambda_{0}$ is
small, the time required for SNs to generate state updates is longer,
resulting in a larger AoI. As $\lambda_{0}$ increases, the time for
SNs to generate state updates decreases, leading to a decrease in
AoI. However, as $\lambda_{0}$ continues to increase, the time for
SNs to generate state updates further decreases, but the UAV's data
collection capacity is limited, resulting in a stabilized AoI. In
Fig. \ref{fig:sample-rate}(b), we present the relationship between
$\lambda_{0}$ and the average energy consumption. Fig. \ref{fig:sample-rate}(b)
reveals that the impact of $\lambda_{0}$ on average energy consumption
is minimal. This is due to the fact that $\lambda_{0}$ primarily
influences the UAV's communication energy consumption, which is considerably
smaller in magnitude when contrasted with the energy consumption during
flight. Fig. \ref{fig:sample-rate}(c) displays the relationship between
the weighted sum and $\lambda_{0}$. Our suggested CADRL-based method
achieves the minimum weighted sum, which demonstrates its superior
performance, as shown in Fig. \ref{fig:sample-rate}(c).

\subsection{The Performance of MLCADRL-based algorithm}

\begin{figure}[tb]
\includegraphics[width=0.45\textwidth]{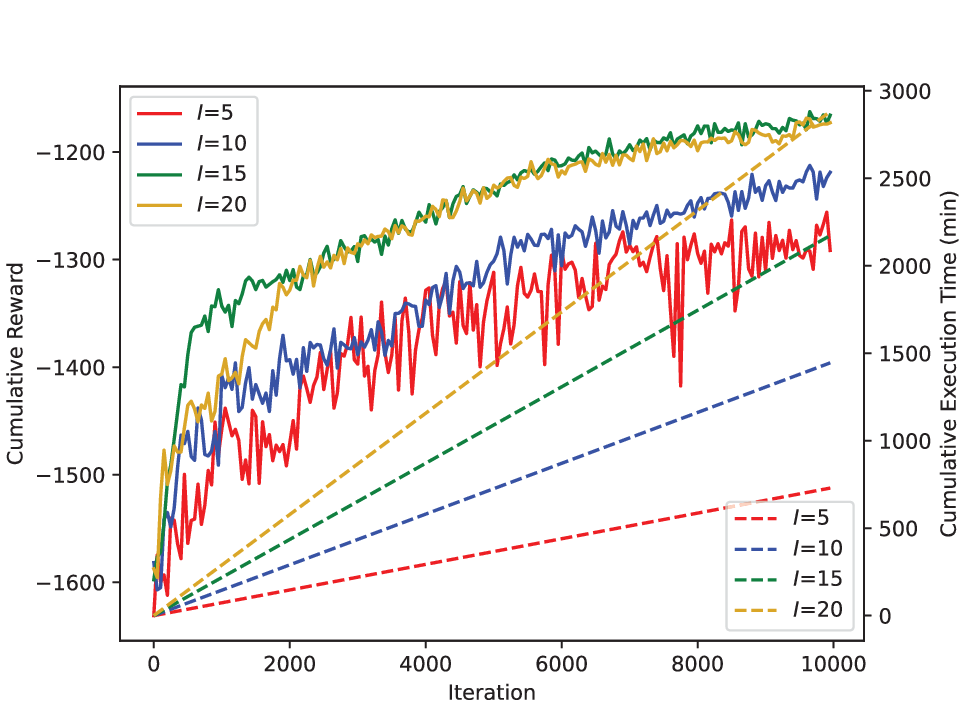}\centering\caption{The influence of the number of tasks selected from the task set $I$
on the convergence performance and training time of the MLCADRL-based
algorithm $(\omega_{1}:\omega_{2}=1:10,\lambda_{0}=0.1\textrm{ and }\lambda_{1}=0.7)$.
\label{fig:MLCADRL-converge-time}}
\end{figure}

Fig. \ref{fig:MLCADRL-converge-time} illustrates the influence of
the number of tasks selected from the task distribution $I$ on the
convergence performance and training time of the MLCADRL-based algorithm
during the training phase. The solid line represents the convergence
performance, while the dashed line represents the convergence time.
From Fig. \ref{fig:MLCADRL-converge-time}, we can observe that as
$I$ increases, the performance of the MLCADRL-based algorithm continuously
improves, and it reaches its best performance when $I=15$. Beyond
this point, further increasing $I$ does not lead to any significant
performance improvement. This is because with more training tasks,
the algorithm gains more knowledge, leading to performance enhancement.
However, when $I$ reaches $15$, the algorithm will have gained a
comprehensive amount of knowledge. On the other hand, as $I$ increases,
the training time of the MLCADRL-based algorithm also increases. This
is because more training tasks require more computational resources,
resulting in a longer training time.

\begin{figure}[tb]
\includegraphics[width=0.4\textwidth]{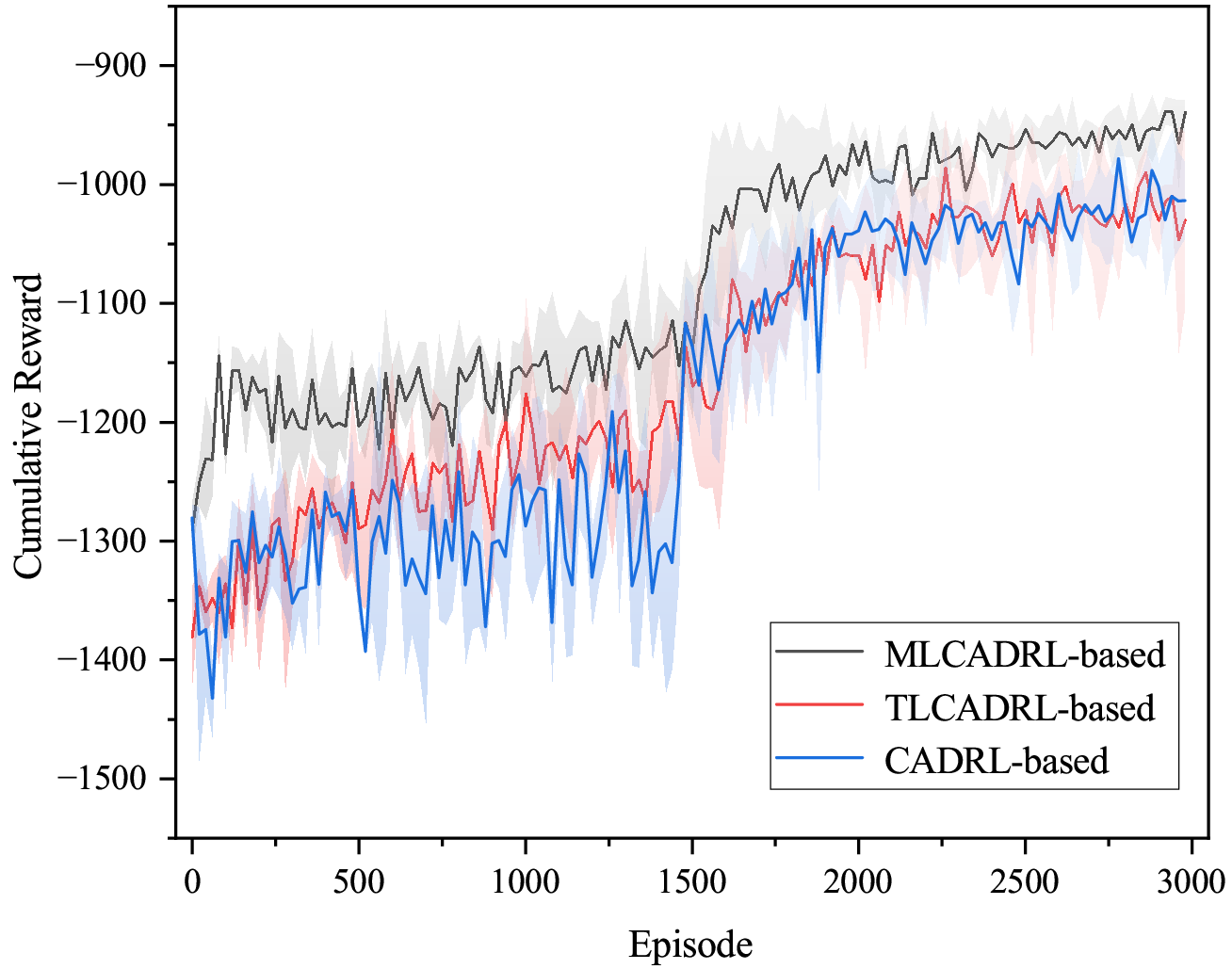}\centering\caption{Performance comparison of algorithms when the task changes. \label{fig:meta-scale}}
\end{figure}

Fig. \ref{fig:meta-scale} illustrates the performance comparison
of three algorithms, namely MLCADRL-based, TLCADRL-based, and CADRL-based
algorithms, while the task changes. At the $1500$th episode, the
number of SNs in the environment decreased from $16$ to $12$, and
the positions of SNs were also altered. From Fig. \ref{fig:meta-scale},
we observe that the MLCADRL-based algorithm adapts more rapidly to
the new task compared to the SLCADRL-based and CADRL-based algorithms.
This can be attributed to the meta-policy learned during the training
phase of the MLCADRL-based algorithm, which enables quick adaptation
to varying environmental conditions. 

\section{Conclusions\label{sec:Conclusions}}

This paper has investigated the problem of timely and energy-efficient
data collection in solar-powered UAV-assisted IoT networks where the
UAV scheduled SNs to gather their status updates, temporarily stored
data packets in its buffer, and subsequently, following an offloading
strategy, offloaded the data packets from its buffers to the DC. Each
SN sampled the environment stochastically. The joint optimization
of the UAV's trajectory, scheduling of SNs, and offloading strategy
was undertaken to minimize the weighted sum of the average AoI of
SNs and energy consumption. The problem has been formulated as a finite-horizon
MDP, and then the CADRL-based and MLCADRL-based data collection algorithms
have been proposed. The CADRL-based algorithm was capable of handling
both continuous and discrete actions of the UAV, and the MLCADRL-based
algorithm, which combines meta-learning and the CADRL-based algorithm,
could improve the performance of CADRL-based algorithms in new tasks.
Simulation results have affirmed the effectiveness of our proposed
algorithms in reducing the weighted sum of the average AoI of SNs
and energy consumption of the UAV compared to the baseline algorithms,
and the combination of meta-learning and CADRL allowed the algorithm
to have fast adaptability to new tasks.

\appendices{}

\bibliographystyle{IEEEtran}
\bibliography{mata_UAV}

\begin{thebibliography}{10}
\providecommand{\url}[1]{#1}
\csname url@samestyle\endcsname
\providecommand{\newblock}{\relax}
\providecommand{\bibinfo}[2]{#2}
\providecommand{\BIBentrySTDinterwordspacing}{\spaceskip=0pt\relax}
\providecommand{\BIBentryALTinterwordstretchfactor}{4}
\providecommand{\BIBentryALTinterwordspacing}{\spaceskip=\fontdimen2\font plus
\BIBentryALTinterwordstretchfactor\fontdimen3\font minus
  \fontdimen4\font\relax}
\providecommand{\BIBforeignlanguage}[2]{{%
\expandafter\ifx\csname l@#1\endcsname\relax
\typeout{** WARNING: IEEEtran.bst: No hyphenation pattern has been}%
\typeout{** loaded for the language `#1'. Using the pattern for}%
\typeout{** the default language instead.}%
\else
\language=\csname l@#1\endcsname
\fi
#2}}
\providecommand{\BIBdecl}{\relax}
\BIBdecl

\bibitem{AI-Fuqaha_IoT-survey_2015IEEECST}
A.~Al-Fuqaha, M.~Guizani, M.~Mohammadi, M.~Aledhari, and M.~Ayyash, ``{Internet
  of Things: A Survey on Enabling Technologies, Protocols, and Applications},''
  \emph{IEEE Commun. Surveys Tuts.}, vol.~17, no.~4, pp. 2347--2376, 2015.

\bibitem{Hossein_UAV-IoT-services-survey_IEEEIoT2016}
N.~Hossein~Motlagh, T.~Taleb, and O.~Arouk, ``{Low-Altitude Unmanned Aerial
  Vehicles-Based Internet of Things Services: Comprehensive Survey and Future
  Perspectives},'' \emph{IEEE Internet Things J.}, vol.~3, no.~6, pp. 899--922,
  2016.

\bibitem{Wu_commu-traj-design-UAV-wireless-net_IEEEWC_2019}
Q.~Wu, L.~Liu, and R.~Zhang, ``{Fundamental Trade-offs in Communication and
  Trajectory Design for UAV-Enabled Wireless Network},'' \emph{IEEE Wireless
  Commun.}, vol.~26, no.~1, pp. 36--44, 2019.

\bibitem{Yang_energy-tradeoff-G2A-traj-design_IEEETVT2018}
D.~Yang, Q.~Wu, Y.~Zeng, and R.~Zhang, ``{Energy Tradeoff in Ground-to-UAV
  Communication via Trajectory Design},'' \emph{IEEE Trans. Veh. Technol.},
  vol.~67, no.~7, pp. 6721--6726, 2018.

\bibitem{Abedin_Data-fresh-EE-drl_IEEETITS2021}
S.~F. Abedin, M.~S. Munir, N.~H. Tran, Z.~Han, and C.~S. Hong, ``{Data
  Freshness and Energy-Efficient UAV Navigation Optimization: A Deep
  Reinforcement Learning Approach},'' \emph{IEEE Trans. Intell. Transp. Syst.},
  vol.~22, no.~9, pp. 5994--6006, 2021.

\bibitem{RDing_3DTra-Freq-EE-FairC-DRL_2020TWC}
R.~{Ding}, F.~{Gao}, and X.~S. {Shen}, ``{3D UAV Trajectory Design and
  Frequency Band Allocation for Energy-Efficient and Fair Communication: A Deep
  Reinforcement Learning Approach},'' \emph{IEEE Trans. Wireless Commun.},
  vol.~19, no.~12, pp. 7796--7809, 2020.

\bibitem{Sun_Solar-power-UAV_TCOM2019}
Y.~Sun, D.~Xu, D.~W.~K. Ng, L.~Dai, and R.~Schober, ``{Optimal 3D-Trajectory
  Design and Resource Allocation for Solar-Powered UAV Communication
  Systems},'' \emph{IEEE Trans. Commun.}, vol.~67, no.~6, pp. 4281--4298, 2019.

\bibitem{Fu_Solar-power-uav_TVT2021}
Y.~Fu, H.~Mei, K.~Wang, and K.~Yang, ``{Joint Optimization of 3D Trajectory and
  Scheduling for Solar-Powered UAV Systems},'' \emph{IEEE Transa. Veh.
  Technol.}, vol.~70, no.~4, pp. 3972--3977, 2021.

\bibitem{UAV_BS_wirelessCharge2018}
A.~{Trotta}, M.~D. {Felice}, F.~{Montori}, K.~R. {Chowdhury}, and L.~{Bononi},
  ``{Joint Coverage, Connectivity, and Charging Strategies for Distributed UAV
  Networks},'' \emph{IEEE Trans. Robot.}, vol.~34, no.~4, pp. 883--900, 2018.

\bibitem{Qian_UAV-MEC_2023IEEEIoTJ}
L.~P. Qian, H.~Zhang, Q.~Wang, Y.~Wu, and B.~Lin, ``{Joint Multi-Domain
  Resource Allocation and Trajectory Optimization in UAV-Assisted Maritime IoT
  Networks},'' \emph{IEEE Internet Things J.}, vol.~10, no.~1, pp. 539--552,
  2023.

\bibitem{Wang_Coverage-UAVs_IEEETWC2022}
L.~Wang, H.~Zhang, S.~Guo, and D.~Yuan, ``{Deployment and Association of
  Multiple UAVs in UAV-Assisted Cellular Networks With the Knowledge of
  Statistical User Position},'' \emph{IEEE Trans. Wireless Commun.}, vol.~21,
  no.~8, pp. 6553--6567, 2022.

\bibitem{Zhang_UAVs-BSs-IEEETWC2023}
X.~Zhang, H.~Zhao, J.~Wei, C.~Yan, J.~Xiong, and X.~Liu, ``{Cooperative
  Trajectory Design of Multiple UAV Base Stations With Heterogeneous Graph
  Neural Networks},'' \emph{IEEE Trans. Wireless Commun.}, vol.~22, no.~3, pp.
  1495--1509, 2023.

\bibitem{S.Kaul_Mini_AoI_VehicularNet}
S.~{Kaul}, M.~{Gruteser}, V.~{Rai}, and J.~{Kenney}, ``{Minimizing Age of
  Information in Vehicular Networks},'' in \emph{Proc. IEEE 8th Annu. Commun.
  Soc. Conf. Sensor, Mesh, Ad Hoc Commun. Netw.}, 2011, pp. 350--358.

\bibitem{WXJ_AoI-Vehicular_IEEENetwork2022}
C.~Guo, X.~Wang, L.~Liang, and G.~Y. Li, ``{Age of Information, Latency, and
  Reliability in Intelligent Vehicular Networks},'' \emph{IEEE Network}, pp.
  1--8, 2022.

\bibitem{abd-elmagidDeepReinforcementLearning2019}
M.~A. {Abd-Elmagid}, A.~Ferdowsi, H.~S. Dhillon, and W.~Saad,
  ``\BIBforeignlanguage{en}{{Deep Reinforcement Learning for Minimizing
  Age-of-Information in UAV-Assisted Networks}},'' in
  \emph{\BIBforeignlanguage{en}{Proc. IEEE Global Commun. Conf. (GLOBECOM)}},
  {Puako, HI, USA}, May 2019.

\bibitem{Zhang_traj-UAVs-BSs-GNN-MARL_2023TWC}
X.~Zhang, H.~Zhao, J.~Wei, C.~Yan, J.~Xiong, and X.~Liu, ``{Cooperative
  Trajectory Design of Multiple UAV Base Stations With Heterogeneous Graph
  Neural Networks},'' \emph{IEEE Trans. Wireless Commun.}, vol.~22, no.~3, pp.
  1495--1509, 2023.

\bibitem{Chu_SpeedControlEnergyReplenishSTDRL_2022IoT}
N.~H. Chu, D.~T. Hoang, D.~N. Nguyen, N.~Van~Huynh, and E.~Dutkiewicz, ``{Joint
  Speed Control and Energy Replenishment Optimization for UAV-assisted IoT Data
  Collection with Deep Reinforcement Transfer Learning},'' \emph{IEEE Internet
  Things J.}, pp. 1--1, 2022.

\bibitem{Zhu_UAV-traj-AoI-Transformer_TWC2023}
B.~Zhu, E.~Bedeer, H.~H. Nguyen, R.~Barton, and Z.~Gao, ``{UAV Trajectory
  Planning for AoI-Minimal Data Collection in UAV-Aided IoT Networks by
  Transformer},'' \emph{IEEE Trans. Wireless Commun.}, vol.~22, no.~2, pp.
  1343--1358, 2023.

\bibitem{JLiu_UAV_AoIWSN}
J.~Liu, P.~Tong, X.~Wang, B.~Bai, and H.~Dai, ``{UAV-Aided Data Collection for
  Information Freshness in Wireless Sensor Networks},'' \emph{IEEE Trans.
  Wireless Commun.}, vol.~20, no.~4, pp. 2368--2382, 2021.

\bibitem{Liu_UAV-traj-AoI-const-env-monitor_IoTJ2022}
K.~Liu and J.~Zheng, ``{UAV Trajectory Optimization for Time-Constrained Data
  Collection in UAV-Enabled Environmental Monitoring Systems},'' \emph{IEEE
  Internet Things J.}, vol.~9, no.~23, pp. 24\,300--24\,314, 2022.

\bibitem{Long_AoIUAVs_2022GLOBECOM}
Y.~Long, W.~Zhang, S.~Gong, X.~Luo, and D.~Niyato, ``{AoI-aware Scheduling and
  Trajectory Optimization for Multi-UAV-assisted Wireless Networks},'' in
  \emph{Proc. IEEE Global Commun. Conf. (GLOBECOM)}, 2022, pp. 2163--2168.

\bibitem{Zhang_AoI-UAV-mURLLC-FBC_JSAC2021}
X.~Zhang, J.~Wang, and H.~V. Poor, ``{AoI-Driven Statistical Delay and
  Error-Rate Bounded QoS Provisioning for mURLLC Over UAV-Multimedia 6G Mobile
  Networks Using FBC},'' \emph{IEEE J. Sel. Areas Commun.}, vol.~39, no.~11,
  pp. 3425--3443, 2021.

\bibitem{WXJ_MultiUAVs-AoI_TCOM2023}
X.~Wang, M.~Yi, J.~Liu, Y.~Zhang, M.~Wang, and B.~Bai, ``{Cooperative Data
  Collection With Multiple UAVs for Information Freshness in the Internet of
  Things},'' \emph{IEEE Trans. Commun.}, vol.~71, no.~5, pp. 2740--2755, 2023.

\bibitem{Liu_AoI-task-assign-traj-multiUAVs_IoTJ2022}
C.~Liu, Y.~Guo, N.~Li, and X.~Song, ``{AoI-Minimal Task Assignment and
  Trajectory Optimization in Multi-UAV-Assisted IoT Networks},'' \emph{IEEE
  Internet Things J.}, vol.~9, no.~21, pp. 21\,777--21\,791, 2022.

\bibitem{Sun_how-keep-data-fresh_IEEETIT2017}
Y.~Sun, E.~Uysal-Biyikoglu, R.~D. Yates, C.~E. Koksal, and N.~B. Shroff,
  ``{Update or Wait: How to Keep Your Data Fresh},'' \emph{IEEE Trans. Inf.
  Theory}, vol.~63, no.~11, pp. 7492--7508, 2017.

\bibitem{MAAE_DRL_AoI_RFPower_2020}
M.~A. {Abd-Elmagid}, H.~S. {Dhillon}, and N.~{Pappas}, ``{A Reinforcement
  Learning Framework for Optimizing Age of Information in RF-Powered
  Communication Systems},'' \emph{IEEE Trans. Commun.}, vol.~68, no.~8, pp.
  4747--4760, 2020.

\bibitem{Zhou_DRL-AoI-UAV-random-sample_WCSP2019}
C.~Zhou, H.~He, P.~Yang, F.~Lyu, W.~Wu, N.~Cheng, and X.~Shen, ``{Deep RL-based
  Trajectory Planning for AoI Minimization in UAV-assisted IoT},'' in
  \emph{Proc. 11th Int. Conf. Wireless Commun. Signal Process. (WCSP), Xian,
  China, Oct.}, 2019, pp. 1--6.

\bibitem{Tong_DRL-EE-AoI-uav_ICC2020}
P.~Tong, J.~Liu, X.~Wang, B.~Bai, and H.~Dai, ``{Deep Reinforcement Learning
  for Efficient Data Collection in UAV-Aided Internet of Things},'' in
  \emph{Proc. IEEE Int. Conf. Commun. Workshops}, 2020, pp. 1--6.

\bibitem{Li_VDN-uav-aoi_IoTJ2023}
Z.~Li, P.~Tong, J.~Liu, X.~Wang, L.~Xie, and H.~Dai, ``{Learning-Based Data
  Gathering for Information Freshness in UAV-Assisted IoT Networks},''
  \emph{IEEE Internet Things J.}, vol.~10, no.~3, pp. 2557--2573, 2023.

\bibitem{Sam_CDRL-solarUAV-noma_IEEEJSAC2021}
S.~Khairy, P.~Balaprakash, L.~X. Cai, and Y.~Cheng, ``{Constrained Deep
  Reinforcement Learning for Energy Sustainable Multi-UAV Based Random Access
  IoT Networks With NOMA},'' \emph{IEEE J. Sel. Areas Commun.}, vol.~39, no.~4,
  pp. 1101--1115, 2021.

\bibitem{Zhang_DRL-hybrid-powered-uav_IEEEIoTJ2023}
Z.~Zhang, C.~Xu, Z.~Li, X.~Zhao, and R.~Wu, ``{Deep Reinforcement Learning for
  Aerial Data Collection in Hybrid-Powered NOMA-IoT Networks},'' \emph{IEEE
  Internet Things J.}, vol.~10, no.~2, pp. 1761--1774, 2023.

\bibitem{Zhang_EE-solar-powered-UAV_IEEETMC2022}
L.~Zhang, A.~Celik, S.~Dang, and B.~Shihada, ``{Energy-Efficient Trajectory
  Optimization for UAV-Assisted IoT Networks},'' \emph{IEEE Trans. Mobile
  Comput.}, vol.~21, no.~12, pp. 4323--4337, 2022.

\bibitem{silver2018general}
D.~Silver, T.~Hubert, J.~Schrittwieser, I.~Antonoglou, M.~Lai, A.~Guez,
  M.~Lanctot, L.~Sifre, D.~Kumaran, T.~Graepel \emph{et~al.}, ``{A General
  Reinforcement Learning Algorithm that Masters Chess, Shogi, and Go Through
  Self-play},'' \emph{Science}, vol. 362, no. 6419, pp. 1140--1144, 2018.

\bibitem{berner2019dota}
C.~Berner, G.~Brockman, B.~Chan, V.~Cheung, P.~D{\k{e}}biak, C.~Dennison,
  D.~Farhi, Q.~Fischer, S.~Hashme, C.~Hesse \emph{et~al.}, ``{Dota 2 with Large
  Scale Deep Reinforcement Learning},'' \emph{arXiv preprint arXiv:1912.06680},
  2019.

\bibitem{rajeswaran2017learning}
A.~Rajeswaran, V.~Kumar, A.~Gupta, G.~Vezzani, J.~Schulman, E.~Todorov, and
  S.~Levine, ``{Learning Complex Dexterous Manipulation with Deep Reinforcement
  Learning and Demonstrations},'' \emph{arXiv preprint arXiv:1709.10087}, 2017.

\bibitem{Yuan_AC-energy-mini-UAV_TVT2021}
Y.~Yuan, L.~Lei, T.~X. Vu, S.~Chatzinotas, S.~Sun, and B.~Ottersten, ``{Energy
  Minimization in UAV-Aided Networks: Actor-Critic Learning for Constrained
  Scheduling Optimization},'' \emph{IEEE Trans. Veh. Technol.}, vol.~70, no.~5,
  pp. 5028--5042, 2021.

\bibitem{Liu_RL-Mul-UAV-deploy-move_TVT2019}
X.~Liu, Y.~Liu, and Y.~Chen, ``{Reinforcement Learning in Multiple-UAV
  Networks: Deployment and Movement Design},'' \emph{IEEE Trans. Veh.
  Technol.}, vol.~68, no.~8, pp. 8036--8049, 2019.

\bibitem{Seid_blockchain-energy-harvesting-UAVs-MADRL_JSAC2022}
A.~M. Seid, J.~Lu, H.~N. Abishu, and T.~A. Ayall, ``{Blockchain-Enabled Task
  Offloading With Energy Harvesting in Multi-UAV-Assisted IoT Networks: A
  Multi-Agent DRL Approach},'' \emph{IEEE J. Sel. Areas Commun.}, vol.~40,
  no.~12, pp. 3517--3532, 2022.

\bibitem{Sun_AoI-TD3-UAV_IoTJ2021}
M.~Sun, X.~Xu, X.~Qin, and P.~Zhang, ``{AoI-Energy-Aware UAV-Assisted Data
  Collection for IoT Networks: A Deep Reinforcement Learning Method},''
  \emph{IEEE Internet Things J.}, vol.~8, no.~24, pp. 17\,275--17\,289, 2021.

\bibitem{LK_JointFlightCruiseControlDataCollectUAVaidedIoT:DRL_IoT2021}
K.~Li, W.~Ni, E.~Tovar, and M.~Guizani, ``{Joint Flight Cruise Control and Data
  Collection in UAV-Aided Internet of Things: An Onboard Deep Reinforcement
  Learning Approach},'' \emph{IEEE Internet Things J.}, vol.~8, no.~12, pp.
  9787--9799, 2021.

\bibitem{Hu_CooperativeUAV-MADRL-CA2C_TCOM2020}
J.~Hu, H.~Zhang, L.~Song, R.~Schober, and H.~V. Poor, ``{Cooperative Internet
  of UAVs: Distributed Trajectory Design by Multi-Agent Deep Reinforcement
  Learning},'' \emph{IEEE Trans. Commun.}, vol.~68, no.~11, pp. 6807--6821,
  2020.

\bibitem{Akbari_AoI-VNF-compound-action_JSAC2021}
M.~Akbari, M.~R. Abedi, R.~Joda, M.~Pourghasemian, N.~Mokari, and
  M.~Erol-Kantarci, ``{Age of Information Aware VNF Scheduling in Industrial
  IoT Using Deep Reinforcement Learning},'' \emph{IEEE J. Sel. Areas Commun.},
  vol.~39, no.~8, pp. 2487--2500, 2021.

\bibitem{fan2019hybrid}
Z.~Fan, R.~Su, W.~Zhang, and Y.~Yu, ``{Hybrid Actor-critic Reinforcement
  Learning in Parameterized Action Space},'' \emph{arXiv preprint
  arXiv:1903.01344}, 2019.

\bibitem{Zhu_UAV-data-collection-S2S-DRL-IEEEIoTJ2022}
B.~Zhu, E.~Bedeer, H.~H. Nguyen, R.~Barton, and J.~Henry, ``{Joint Cluster Head
  Selection and Trajectory Planning in UAV-Aided IoT Networks by Reinforcement
  Learning With Sequential Model},'' \emph{IEEE Internet Things J.}, vol.~9,
  no.~14, pp. 12\,071--12\,084, 2022.

\bibitem{Yi_Multi-task-DRL-UAV-IEEEIoTJ2023}
M.~Yi, X.~Wang, J.~Liu, Y.~Zhang, and R.~Hou, ``{Multi-Task Transfer Deep
  Reinforcement Learning for Timely Data Collection in Rechargeable-UAV-aided
  IoT Networks},'' \emph{IEEE Internet Things J.}, pp. 1--1, 2023.

\bibitem{Lu_UAV-metaDRL-traj_INFOCOM2023}
Z.~Lu, X.~Wang, and M.~C. Gursoy, ``{Trajectory Design for Unmanned Aerial
  Vehicles via Meta-Reinforcement Learning},'' in \emph{Proc. IEEE Int. Conf.
  Comput.Commun. Workshops (INFOCOM WKSHPS)}, 2023, pp. 1--6.

\bibitem{Samir_large&small_scale_TWC2020}
M.~Samir, S.~Sharafeddine, C.~M. Assi, T.~M. Nguyen, and A.~Ghrayeb, ``{UAV
  Trajectory Planning for Data Collection from Time-Constrained IoT Devices},''
  \emph{IEEE Trans. Wireless Commun.}, vol.~19, no.~1, pp. 34--46, 2020.

\bibitem{Tran_large&small_scale_TWC2022}
D.-H. Tran, V.-D. Nguyen, S.~Chatzinotas, T.~X. Vu, and B.~Ottersten, ``{UAV
  Relay-Assisted Emergency Communications in IoT Networks: Resource Allocation
  and Trajectory Optimization},'' \emph{IEEE Trans. Wireless Commun.}, vol.~21,
  no.~3, pp. 1621--1637, 2022.

\bibitem{al2014optimal}
A.~Al-Hourani, S.~Kandeepan, and S.~Lardner, ``{Optimal LAP Altitude for
  Maximum Coverage},'' \emph{IEEE Wireless Commun. Lett.}, vol.~3, no.~6, pp.
  569--572, 2014.

\bibitem{sn-energy-bonuli1_JSAC2016}
A.~Baknina and S.~Ulukus, ``{Optimal and Near-Optimal Online Strategies for
  Energy Harvesting Broadcast Channels},'' \emph{IEEE J. Sel. Areas Commun.},
  vol.~34, no.~12, pp. 3696--3708, 2016.

\bibitem{schulman2015TRPO}
J.~Schulman, S.~Levine, P.~Abbeel, M.~Jordan, and P.~Moritz, ``{Trust region
  policy optimization},'' in \emph{Proc. Int. Conf. Mach. Learn.}\hskip 1em
  plus 0.5em minus 0.4em\relax PMLR, 2015, pp. 1889--1897.

\bibitem{schulman2015-GAE}
J.~Schulman, P.~Moritz, S.~Levine, M.~Jordan, and P.~Abbeel,
  ``{High-dimensional Continuous Control Using Generalized Advantage
  Estimation},'' \emph{arXiv preprint arXiv:1506.02438}, 2015.

\bibitem{thrun1998learning}
S.~Thrun and L.~Pratt, ``{Learning to Learn: Introduction and Overview},'' in
  \emph{Learning to Learn}.\hskip 1em plus 0.5em minus 0.4em\relax Springer,
  1998, pp. 3--17.

\bibitem{finn2017model}
C.~Finn, P.~Abbeel, and S.~Levine, ``{Model-agnostic Meta-learning for Fast
  Adaptation of Deep Networks},'' in \emph{Proc. Int. Conf.Mach. Learn.}\hskip
  1em plus 0.5em minus 0.4em\relax PMLR, 2017, pp. 1126--1135.

\bibitem{williams1992simple}
R.~J. Williams, ``{Simple Statistical Gradient-following Algorithms for
  Connectionist Reinforcement Learning},'' \emph{Mach. Learn.}, pp. 5--32,
  1992.

\bibitem{DQN}
V.~Mnih, K.~Kavukcuoglu, D.~Silver, A.~A. Rusu, J.~Veness, M.~G. Bellemare,
  A.~Graves, M.~Riedmiller, A.~K. Fidjeland, G.~Ostrovski \emph{et~al.},
  ``{Human-level Control Through Deep Reinforcement Learning},'' \emph{Nature},
  vol. 518, no. 7540, pp. 529--533, 2015.

\bibitem{DDPG}
T.~P. Lillicrap, J.~J. Hunt, A.~Pritzel, N.~Heess, T.~Erez, Y.~Tassa,
  D.~Silver, and D.~Wierstra, ``{Continuous Control with Deep Reinforcement
  Learning},'' \emph{arXiv preprint arXiv:1509.02971}, 2015.

\bibitem{xiong2018parametrized}
J.~Xiong, Q.~Wang, Z.~Yang, P.~Sun, L.~Han, Y.~Zheng, H.~Fu, T.~Zhang, J.~Liu,
  and H.~Liu, ``{Parametrized Deep Q-networks Learning: Reinforcement Learning
  with Discrete-continuous Hybrid Action Space},'' \emph{arXiv preprint
  arXiv:1810.06394}, 2018.

\bibitem{taylor2009transfer}
M.~E. Taylor and P.~Stone, ``{Transfer Learning for Reinforcement Learning
  Domains: A Survey.}'' \emph{J. Mach. Learn. Res.}, vol.~10, no.~7, 2009.

\end{thebibliography}

\end{document}